\documentclass[10pt,journal,compsoc]{IEEEtran}

\ifCLASSOPTIONcompsoc
  \usepackage[nocompress]{cite}
\else
  \usepackage{cite}
\fi

\usepackage{hyperref}
\usepackage{multirow}
\usepackage[pdftex]{graphicx}
\graphicspath{{/images/}}
\DeclareGraphicsExtensions{.pdf}

\usepackage{amsmath,amssymb,amsfonts}

\usepackage{algorithmic}

\usepackage{array}

\ifCLASSOPTIONcompsoc
  \usepackage[caption=false,font=footnotesize,labelfont=sf,textfont=sf]{subfig}
\else
  \usepackage[caption=false,font=footnotesize]{subfig}
\fi

\usepackage{url}

\usepackage{color}

\usepackage{xcolor,colortbl}
\xdefinecolor{gray95}{gray}{0.65}
\usepackage{comment}

\usepackage{color}
\definecolor{pblue}{rgb}{0.13,0.13,1}
\definecolor{pgreen}{rgb}{0,0.5,0}
\definecolor{pred}{rgb}{0.9,0,0}
\definecolor{pgrey}{rgb}{0.46,0.45,0.48}

\usepackage{listings}
\newcommand{\etal}{\textit{et al. }}

\begin{document}

\title{Automatizing Software Cognitive Complexity Reduction through Integer Linear Programming}

\author{Rubén Saborido,
        Javier Ferrer,
        and~Francisco~Chicano
\IEEEcompsocitemizethanks{\IEEEcompsocthanksitem Authors are with ITIS Software, Universidad de Málaga, Spain.\protect\\
E-mails: rsain@uma.es, jferrer@uma.es, and chicano@uma.es}% <-this % stops an unwanted space
\thanks{Preprint 2023.}}

\markboth{}%
{Saborido \MakeLowercase{\textit{et al.}}: Automatizing Software Cognitive Complexity Reduction through Integer Linear Programming}

\IEEEtitleabstractindextext{%
\begin{abstract}
Reducing the cognitive complexity of a piece of code to a given threshold is not trivial. Recently, we modeled software cognitive complexity reduction as an optimization problem and we proposed an approach to assist developers on this task. This approach enumerates sequences of code extraction refactoring operations until a stopping criteria is met. As a result, it returns the minimal sequence of code extraction refactoring operations that is able to reduce the cognitive complexity of a code to the given threshold. However, exhaustive enumeration algorithms fail to scale with the code size. The number of refactoring plans can grow exponentially with the number of lines of code. In this paper, instead of enumerating sequences of code extraction refactoring operations, we model the cognitive complexity reduction as an Integer Linear Programming problem. This opens the door to the use of efficient solvers to find optimal solutions in large programs.
\end{abstract}

}

\maketitle

\IEEEdisplaynontitleabstractindextext

\IEEEpeerreviewmaketitle

\IEEEraisesectionheading{\section{Introduction}\label{sec:introduction}}
\IEEEPARstart{R}{ecently}, a novel cognitive complexity metric has been proposed and integrated in the well-known static code tools SonarCloud\footnote{\url{https://sonarcloud.io}} and SonarQube\footnote{\url{https://www.sonarqube.org}}, an open-source service and platform, respectively, for continuous inspection of code quality. This cognitive complexity metric, which we refer to as \textbf{SonarSource Cognitive Complexity (SSCC)}, has been defined as a measure of how hard the control flow of a code (function/method) is to understand and maintain~\cite{Campbell2018}. 
The SSCC is given by a positive number which is increased every time a control flow sentence appears. Their nested levels also contribute to the SSCC of the code.
Functions/methods with high SSCC will be difficult to maintain. Although SonarQube suggests to keep code's cognitive complexity no greater than a threshold, software developers lack support to reduce the SSCC of their code.

In a previous work, we modelled the reduction of the SSCC to a given threshold as an optimization problem where the search space contains all feasible sequences of code extraction opportunities \cite{Saborido2022}. We also defined two algorithms which enumerate sequences of code extraction refactoring operations until a stopping criterion is met. As a result, it returns the minimal sequence of code extraction refactoring operations that is able to reduce the cognitive complexity of a code to the given
threshold. However, enumeration algorithms fail to scale with the code size. Thus, the number of refactoring plans can grow exponentially
with the number of lines of code (LOC).

We now introduce a different modeling of the problem, defining the SSCC reduction task as an Integer Linear Programming (ILP) problem. ILP is a type of optimization problem where the variables are integer values and the objective function and equations are linear. This formulation allows us to apply ILP solvers, like CPLEX, to get optimal solutions very quickly. 

We here propose an approach to reduce the SSCC of software projects in an automated way when modeling the cognitive complexity task as an ILP problem. We integrate the proposed approach in a software tool for Java code, and we validate it over 10 open-source software projects to reduce their SSCC. The developed tool will be available as an open-source project in a public repository\footnote{https://github.com/rsain/SoftwareCognitiveComplexityReducer}. Thus, the contributions of this work are threefold:

\begin{itemize}
\item Modeling the SSCC reduction to a given threshold as an ILP problem. 
\item Validating the proposed approach over 10 open-source software projects.
\item Providing a software tool to reduce the SSCC of Java projects in an automated way. 
\end{itemize}

The remainder of this paper is organized as follows. \mbox{Sec. \ref{sec:problem}} motivates the challenge of reducing the SSCC of code and introduces some concepts. \mbox{Sec. \ref{sec:ilp}} presents the novel formulation of SSCC reduction as an ILP  problem. \mbox{Sec. \ref{sec:approach}} introduces our approach for reducing the SSCC to a given
threshold. \mbox{Sec. \ref{sec:casestudy}} presents the case of study and
summarize the experimental setting for evaluating our proposal. \mbox{Sec. \ref{sec:results}} provides the results of our experiments. \mbox{Sec. \ref{sec:discussion}} discusses the benefits of the proposed approach with respect to previous work. \mbox{Sec. \ref{sec:threats}} discusses the threats to the validity of our work. \mbox{Sec. \ref{sec:relatedWork}} summarizes related work. Finally, \mbox{Sec. \ref{sec:conclusion}} presents conclusions and future work.

\section{Motivation and concepts}
\label{sec:problem}
\label{sec:challenge}
In order to illustrate the difficulties developers face when reducing the SSCC of code, we use the \texttt{addParametersToServiceUrl} method in the \texttt{DocumentExecutionWorkForDoc} class of the Knowage-core open-source Java project. Fig.~\ref{fig:ccMotivationBeforeReduction} shows the source code of this method. Note that some code has been replaced by ``...'' due to space limitations, but the whole code is accessible in the following URL\footnote{\url{https://github.com/KnowageLabs/Knowage-Server/blob/dfed28a869125c51e51c66e433acd6e12b359828/knowage-core/src/main/java/it/eng/knowage/api/dossier/DocumentExecutionWorkForDoc.java\#L330}}. 
This method has SSCC 46 and SonarQube suggests reducing it to 15 in order to improve the understandability and maintainability of the method. 
The number of different Extract Method refactoring opportunities of a method with $n$ sentences is bounded by ${n\choose 2} = \frac{n \cdot (n-1)}{2}$. This is the number of combinations of $n$ sentences taken two at a time without repetition. These two sentences determine the beginning and ending of a code extraction.
The \texttt{addParametersToServiceURL} method has 37 sentences and, thus, the upper bound of Extract Method refactoring opportunities is ${37\choose 2}=666$.  

The problem developers face is to find sequences of code extractions that reduce the SSCC of the method. We have computationally checked that there are 31 applicable code extractions which reduce the SSCC of this method. However, one would need to evaluate all possible sequences of Extract Method operations totaling $2^{31}$ alternatives. Note that we use here ``Extract Method" instead of ``Extract Function" to reflect the object-centric nature of the operation in Java. After analyzing the method, a developer who faces this cognitive complexity reduction task could realize that a sequence of at least three Extract Method refactorings is required. Therefore, one would need to evaluate all possible sequences of three Extract Method operations totaling ${31 \choose 3} = 4,495$.

\begin{figure}[!htb]
\centering
\noindent
\begin{lstlisting}[numbers=left,xleftmargin=+2em,frame=single,framexleftmargin=0em,escapechar=ñ,stepnumber=1,
    showstringspaces=false,
    tabsize=1,
    breaklines=true,
    breakatwhitespace=false,basicstyle=\scriptsize,language=Java]{}
public String addParametersToServiceUrl(...) throws ... 
{
 List<BIObjectParameter> drivers = ...;
 if (drivers != null) {
  List<Parameter> parameter = ...;
  if (drivers.size() != parameter.size()) {
    throw new SpagoBIRuntimeException("There are...");
  }
  Collections.sort(drivers);
  ParametersDecoder decoder = new ParametersDecoder();
  for (BIObjectParameter biObjectParameter : drivers) {
   boolean found = false;
   String value = "";
   String paramName = "";
   for (Parameter templateParameter : parameter) {
    if (templateParameter.getType().equals("dynamic") {
     if (templateParameter.getValue() != null && ...) {
      value = templateParameter.getValue();
      if (... && value.contains("STRING"))
       value.replaceAll("'", "");
      if (...) {
       paramName = templateParameter.getUrlName();
       serviceUrlBuilder.append(...);
       serviceUrlBuilder.append(...);
       found = true;
       break;
      }
     }
    } 
    else {
     if (biObjectParameter.getParameterUrlName...) {
      serviceUrlBuilder.append(...);
      value = templateParameter.getValue();
      paramName = templateParameter.getUrlName();
      if (templateParameter.getUrlNameDescription...) {
       throw new SpagoBIRuntimeException("...");
      }
      serviceUrlBuilder.append(...);
      found = true;
      break;
     }
    }
   }
   paramMap.put(paramName, value);
   if (!found) {
    throw new SpagoBIRuntimeException("...");
   }
  }
 }
 return serviceUrlBuilder.toString();
}
\end{lstlisting}
\caption{Method in the \texttt{DocumentExecutionWorkForDoc} class of the Knowage-core software project. The method has SSCC 46 and its reduction to 15 requires to evaluate more than 3,000 Extract Method refactoring opportunities.}
\label{fig:ccMotivationBeforeReduction}
\end{figure}

%\subsection{Concepts}
Next, we introduce several concepts which will be used along the paper.
With the aim of easing the understanding of these concepts, we use as \textit{illustrating example} the \texttt{hook} method in the \texttt{EZInjection} class of the ByteCode open-source java project. Fig. \ref{fig:ccExampleInCode} shows its source code. Note that we added comments to the original source code to show the value of different cognitive complexity metrics of those statements contributing to the SSCC of the method. 

\begin{figure}[!htb]
\begin{lstlisting}[numbers=left,xleftmargin=+0.6cm,xrightmargin=+0.2cm,frame=single,framexleftmargin=0em,escapechar=ñ,stepnumber=1,
    showstringspaces=false,
    tabsize=1,
    breaklines=true,
    breakatwhitespace=false,
    basicstyle=\scriptsize,
    language=Java,
    mathescape=true]{}
//Cognitive complexity 16
public static void hook(String info) {
    // [$\color{pgreen} \lambda$=0, $\color{pgreen} \iota$=1, $\color{pgreen} \nu$=0, $\color{pgreen} \mu$=0, CCR=1, NMCC=1]
    for (BytecodeHook hook : hookArray)
        hook.callHook(info);

    // [$\color{pgreen} \lambda$=0, $\color{pgreen} \iota$=7, $\color{pgreen} \nu$=8, $\color{pgreen} \mu$=5, CCR=15, NMCC=15]
    if (debugHooks) {
        // [$\color{pgreen} \lambda$=1, $\color{pgreen} \iota$=1, $\color{pgreen} \nu$=0, $\color{pgreen} \mu$=1, CCR=2, NMCC=1]
        if (lastMessage.equals(info))
            return;

        lastMessage = info;
        boolean print = all;

        // [$\color{pgreen} \lambda$=1, $\color{pgreen} \iota$=4, $\color{pgreen} \nu$=3, $\color{pgreen} \mu$=3, CCR=10, NMCC=7]
        if (!all && debugClasses.length >= 1) {
            // [$\color{pgreen} \lambda$=2, $\color{pgreen} \iota$=2, $\color{pgreen} \nu$=1, $\color{pgreen} \mu$=2, CCR=7, NMCC=3]
            for (String s : debugClasses) {
                // [$\color{pgreen} \lambda$=3, $\color{pgreen} \iota$=1, $\color{pgreen} \nu$=0, $\color{pgreen} \mu$=1, CCR=4, NMCC=1]
                if (info.split("\\.")[0].equals(s.replaceAll("\\.", "/"))) {
                    print = true;
                    break;
                }
            }
        }

        // [$\color{pgreen} \lambda$=1, $\color{pgreen} \iota$=1, $\color{pgreen} \nu$=0, $\color{pgreen} \mu$=1, CCR=2, NMCC=1]
        if (print)
            print("Method call: " + info);
    }
}
\end{lstlisting}
\caption{Method \texttt{hook} in the \texttt{EZInjection} class of the open-source project ByteCode, used as illustrating example in the paper. We include comments showing cognitive complexity metrics of those statements contributing to the cognitive complexity of the method. $\lambda$, $\iota$, $\nu$, $\mu$, $CCR$, and $NMCC$ refer to the nesting component, accumulated inherent complexity, accumulated nesting complexity, number of nodes contributing to the accumulated nesting complexity, cognitive complexity reduction, and cognitive complexity of the new method when extracted, respectively.}
\label{fig:ccExampleInCode}
\end{figure}

\subsection*{Notation}
We formulated the reduction of SSCC to a given threshold as an optimization problem \cite{Saborido2022}. The SSCC of a piece of code can be computed as the sum of two components: the \textit{inherent component} and the \textit{nesting component}. The inherent component depends on the presence of certain control flow structures and complex expressions (like conditionals combining several kinds of logical operators). Control flow structures and complex expressions contribute +1 to the inherent component. The nesting component, which we refer to $\lambda$, depends on the depth that a certain control flow structure is in the code with respect to the function implementation/method declaration. This depth is the contribution to the nesting component. Note that $\lambda$ is 0 when no nesting exists in the target piece of software.

Let $s_i$ and $e_i$ be the start and end offset (in characters) of the $i$th extractable sequence in the source file. We consider the $i$th sequence is nested in the $j$th  sequence, denoted with $i \rightarrow j$, when $[s_i,e_i] \subset [s_j,e_j]$. Regarding the illustrating example, the second \texttt{for} loop (lines 19-25) is nested in the first conditional (lines 8-31). In order to simplify the notation, we define the code associated to the body of the original function/method as the 0th sequence. Thus, we can write $i \rightarrow 0$ for all $i\geq 1$. We say that the $i$th sequence is \emph{in conflict with} the $j$th sequence, denoted with $i \nleftrightarrow j$ when $i$ and $j$ are not nested one in the other, and $[s_i,e_i] \cap [s_j,e_j] \neq \emptyset$. Two sequences cannot be extracted simultaneously if they are in conflict. In the illustrating example, the sequence defined by lines 10-14 is in conflict with the sequence defined by lines 13-26. Thus, these two sequences cannot be extracted at the same time.

Let $d_{ij}$, for $i \rightarrow j$, be the nesting distance between sequences $i$ and $j$ ($d_{ij} = \lambda{_i} - \lambda{_j}$). That is, the number of %control flow structures increasing the nesting degree
AST nodes increasing the nesting degree of the method when traversing its Abstract Syntax Tree (AST) moving upwards from the AST node corresponding to the $i$ sequence to the AST node of the $j$ sequence. In the illustrating example, the nesting distance between the first conditional (lines 8-31) and the second \texttt{for} loop (lines 19-25) is two. We also denote with $\iota_i$ the \textit{accumulated inherent component} of the $i$th sequence and the ones contained by it. We define $\nu_i$ as the \textit{accumulated  nesting component} of the sequences contained by the $i$th sequence (considering nesting 0 for the $i$th sequence). We denote with $\mu_i$ the number of sequences that contribute to the accumulated nesting component of the $i$th sequence (including itself when its nesting component is not 0). We finally define $CCR_i$ and $NMCC_i$ as the cognitive complexity reduction and cognitive complexity of the new method when extracting sequence $i$, respectively. Note that $CCR_i = (\mu_i * \lambda_i) + \iota_i + \nu_i$ and $NMCC_i = \iota_i + \nu_i$.

\subsection*{Refactoring cache of a method}
We also define a cache of Extract Method refactoring opportunities associated to a method in order to reduce its cognitive complexity, which we name the \emph{refactoring cache} of a method. The refactoring cache contains the start and end offset of the $i$th extraction ($s_i$ and $e_i$, respectively), the feasibility of the extraction, the reason the extraction is not feasible or ``OK" otherwise, the number of parameters of the new method signature, the number of lines of code of the extraction, $CCR_i$, $NMCC_i$, $\iota_i$, $\nu_i$, $\mu_i$, and $\lambda_i$. Note that the source code of a method is not modified when generating its refactoring cache. We use the refactoring cache to speed up the search of sequences of code extractions when enumerating single extract method refactoring operations. Thus, when a code extraction is evaluated, the refactoring cache is queried. If there is a hit, cognitive complexity metrics of this code extraction are obtained from the cache. If not, we (i) evaluate the feasibility of the code extraction, (ii) compute associated metrics, and (iii) update the cache with computed metrics.  Fig.~\ref{fig:refactoringCache} shows the refactoring cache of the illustrating example. Note that the fourth column has been shortened by replacing part of the text with ``..." in order to ease the readability of the figure. As shown, the method used as illustrating example has 17 candidate code extractions involving statements contributing to the overall cognitive complexity of the method. However, six of these extractions are not feasible: (i) five extractions (lines 4-8 in Fig.~\ref{fig:refactoringCache}) include a \texttt{return} statement (line 11 in Fig.~\ref{fig:ccExampleInCode}), but not all possible execution flows would end in a \texttt{return} in the extracted code (semantics may not be preserved) and (ii) one extraction (line 16 in Fig.~\ref{fig:refactoringCache}, corresponding to lines 21-24 in Fig.~\ref{fig:ccExampleInCode}) contains a \texttt{break} statement but not the parent statement it belongs to. Note that the refactoring cache includes a code extraction for the whole body of the method (line 2 in Fig.~\ref{fig:refactoringCache}). Although this extraction is feasible, it would create a new method with the same code of the original method. Therefore, this extraction is not considered when solving the cognitive complexity reduction task. However, we include it in the refactoring cache for completeness.

\begin{figure}[!htb]
\begin{lstlisting}[numbers=left,xleftmargin=+0.6cm,xrightmargin=+0.2cm,frame=single,framexleftmargin=0em,escapechar=ñ,stepnumber=1,
    showstringspaces=false,
    tabsize=1,
    breaklines=true,
    breakatwhitespace=false,
    basicstyle=\scriptsize,
    language=,
    mathescape=true]{}
4695, 4763, 1, "OK", 1, 2, 1, 1, 1, 0, 0, 0
4695, 5402, 1, "OK", 1, 26, 16, 16, 8, 8, 5, 0
4773, 5402, 1, "OK", 1, 23, 15, 15, 7, 8, 5, 0
4811, 4902, 0, "Selected ...", 0, 0, 2, 1, 1, 0, 1, 1
4811, 4935, 0, "Selected ...", 0, 0, 2, 1, 1, 0, 1, 1
4811, 4968, 0, "Selected ...", 0, 0, 2, 1, 1, 0, 1, 1
4811, 5321, 0, "Selected ...", 0, 0, 12, 8, 5, 3, 4, 1
4811, 5392, 0, "Selected ...", 0, 0, 14, 9, 6, 3, 5, 1
4916, 5321, 1, "OK", 1, 14, 10, 7, 4, 3, 3, 1
4916, 5392, 1, "OK", 1, 17, 12, 8, 5, 3, 4, 1
4948, 5321, 1, "OK", 1, 13, 10, 7, 4, 3, 3, 1
4948, 5392, 1, "OK", 1, 16, 12, 8, 5, 3, 4, 1
4982, 5321, 1, "OK", 2, 11, 10, 7, 4, 3, 3, 1
4982, 5392, 1, "OK", 2, 14, 12, 8, 5, 3, 4, 1
5050, 5307, 1, "OK", 2, 8, 7, 3, 2, 1, 2, 2
5118, 5289, 0, "Section ...", 0, 0, 4, 1, 1, 0, 1, 3
5335, 5392, 1, "OK", 2, 2, 2, 1, 1, 0, 1, 1
\end{lstlisting}
\caption{Refactoring cache of the illustrating example. The refactoring cache contains the start and end offset of the $i$th extraction ($s_i$ and $e_i$, respectively), the feasibility of the extraction, the reason the extraction is not feasible or ``OK" otherwise, the number of parameters of the new method signature, the number of lines of code of the extraction, $CCR_i$, $NMCC_i$, $\iota_i$, $\nu_i$, $\mu_i$, and $\lambda_i$.}
\label{fig:refactoringCache}
\end{figure}

\subsection*{Conflict graph of a method}
\begin{figure}[!htb]
    \includegraphics[scale=0.35]{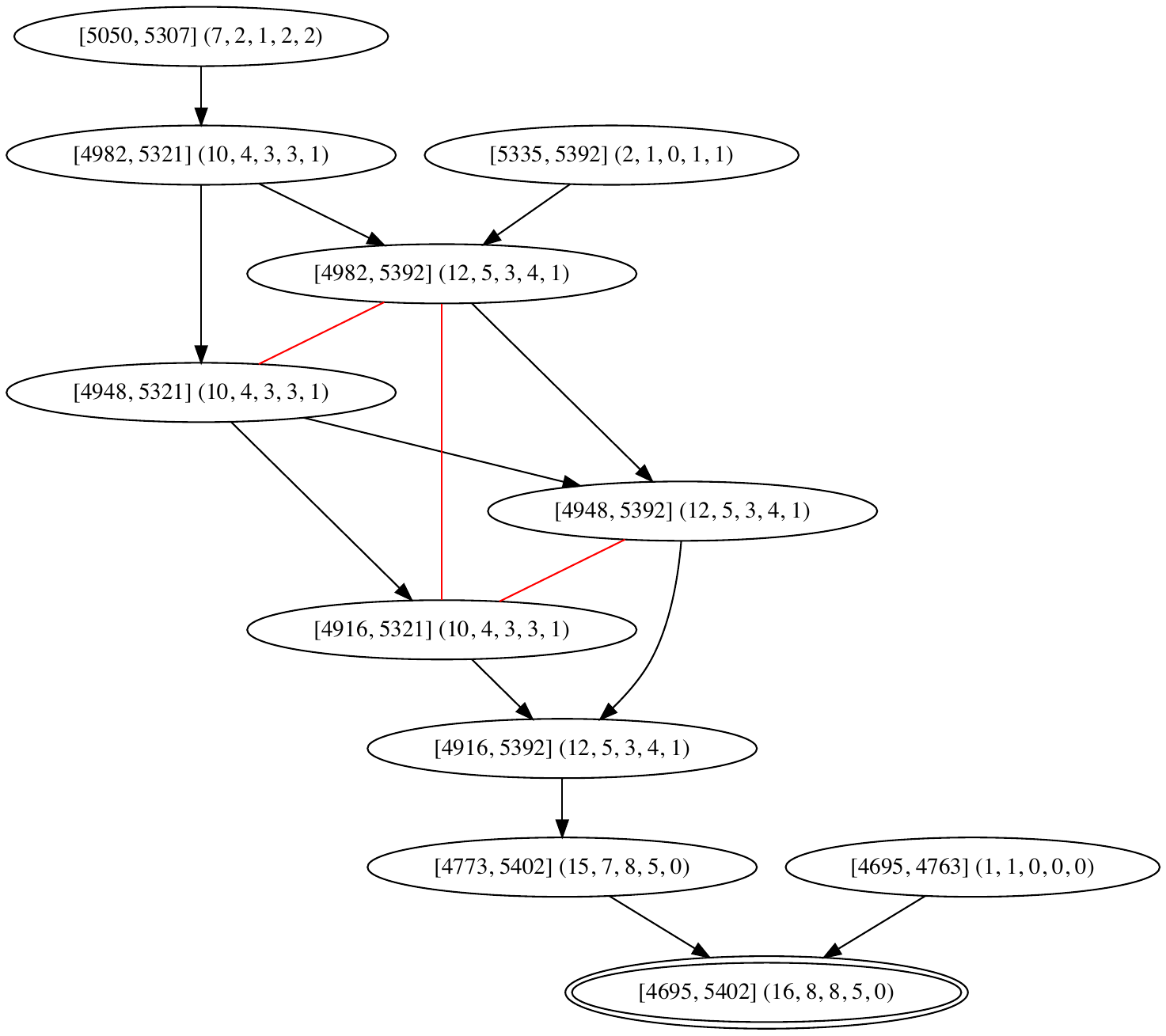}
    \caption{Conflict graph of the illustrating example.}
    \label{fig:graphHook}
\end{figure}

Based on the information provided by the refactoring cache of a method, we generate what we name the \textit{conflict graph} of a method. A conflict graph is a graph where the vertices are applicable extractions and there are two types of edges: \emph{nested} and \emph{conflict} edges. There is a nested edge from node $i$ to node $j$ if $i$ is  a sequence nested in $j$: $i \rightarrow j$. There is a conflict edge between nodes $i$ and $j$ if the sequences are in conflict: $i \nleftrightarrow j$. We label the vertices in the conflict graph as \mbox{$[s_i, e_i] (CCR_i, \iota_i, \nu_i, \mu_i,\lambda_i)$}, where $s_i$ and $e_i$ refer to the start and end offset (in characters in the source file) of the $i$th extraction. We represent the conflict edges with red edges in the representation of the graph. Note that two vertices in conflict cannot be both selected for extraction in the same sequence. The conflict graph also includes a special vertex representing the whole body of the method. This vertex is represented with double ellipse. The conflict graph is useful to have a visual representation of extractable sequences in a method, but it is also useful to compute the impact of code extractions when reducing the SSCC of a method. Fig.~\ref{fig:graphHook} shows the conflict graph of the illustrating example generated from the refactoring cache of the method (\mbox{Fig. \ref{fig:refactoringCache}}). Although the \texttt{hook} method has 17 candidate code extractions, 11 code extractions are feasible. Thus, we can observe in the conflict graph 10 applicable extractions (vertices) plus the one representing the whole body of the method, which is located in the bottom. As shown in the conflict graph, there are 12 nested edges (in black) and three conflict edges (in red).

\section{The ILP problem definition}
\label{sec:ilp}
A zero-one program is an integer program in which the variables can only take values 0 or 1. In this section we describe the zero-one program to model the reduction of the SSCC of a function/method to a given threshold $\tau$:
\begin{small}
\begin{align}
\label{eqn:objective} & \min \sum_{i=1}^m x_i \\
\intertext{subject to:}
\label{eqn:conflict} & x_i + x_j \leq 1 \;\; \forall i \nleftrightarrow j \\
\label{eqn:limit} &  \underbrace{(\iota_i + \nu_i)}_{\text{Term A}}  x_i - \underbrace{\sum_{j, j \rightarrow i} z_{ji} (\iota_j + \nu_j + d_{ji} \mu_j)}_{\text{Term B}}  \leq \tau \;\; \forall i=0,1, \ldots, m\\
\label{eqn:zdef} & z_{ji} + \left|\left\{l\middle|j \rightarrow l \rightarrow i \right\}\right| (z_{ji}-1) \leq x_j - \sum_{l, j \rightarrow l \rightarrow i} x_l   \;\; \forall j \rightarrow i\\
\label{eqn:x0} & x_0 = 1\\
\label{eqn:domain-x}& x_i \in \{0,1\} \;\; \forall i=0, 1, 2, \ldots, m \\
\label{eqn:domain-z}& z_{ji} \in \{0,1\} \;\; \forall j \rightarrow i
\end{align}
\end{small}

\noindent where $m$ is the number of sequences that can be extracted and $x_i$ is a binary variable that takes value 1 if sequence $i$ is extracted and 0 otherwise.
The goal, expressed in Eq.~\eqref{eqn:objective}, is to minimize the number of extracted sequences. 
Eq.~\eqref{eqn:conflict} ensures that at most one of two sequences in conflict will be selected in the final solution. 
Eq.~\eqref{eqn:limit} limits the SSCC of all the sequences in the code after applying the refactoring operations to be lower or equal than $\tau$. In the left-hand side, we find the SSCC of the $i$th sequence when extracted as a new function/method (\text{Term A}) minus SSCC removed from sequence $i$ due to code extractions of sequences $j$ contained in $i$ (\text{Term B}). If there were two sequences $k$ and $j$ with $k \rightarrow j \rightarrow i$ that are extracted, we should only subtract the $j$th sequence SSCC contribution, and not $k$. The reason is that the contribution of sequence $k$ is already considered in the contribution of $j$. For this reason, we do not use the $x$ variables in the expression of the SSCC reduction (\text{Term B}). Instead, we define new variables $z_{ji}$ for $j \rightarrow i$ that are 1 if and only if the $j$th extractable sequence is selected and no other sequence strictly between $j$ and $i$ is selected. If $z_{ji}=1$ then we should extract from the SSCC of the $i$th sequence the SSCC of the $j$th sequence. This is what Eq.~\eqref{eqn:limit} does.
Variables $z$ are completely determined by the values of the $x$ variables. Eq.~\eqref{eqn:zdef} expresses the relationship between the $z$ variables and the $x$ variables. If $z_{ji}=1$, then $x_j=1$ and $x_l=0$ for all $l$ with $j \rightarrow l \rightarrow i$. If $z_{ji}=0$, then there is no constraint for the $x$ variables. Eq.~\eqref{eqn:x0} forces the SSCC reduction of the original function/method (represented with the 0$th$ sequence).
Finally, Eqs.~\eqref{eqn:domain-x} and~\eqref{eqn:domain-z} define the binary domain of the variables.

\subsection*{ILP concepts definition}
In order to solve an ILP problem, solvers usually implement optimizers based on the simplex algorithms (both primal and dual simplex). Usually, the optimizer determines a model as \emph{infeasible}, \emph{feasible}, \emph{unknown} or \emph{optimal}. 
Infeasible specifies that the algorithm proved the model infeasible (that is, it is not possible to find an assignment of values to variables satisfying all the constraints in the model). Feasible specifies that the algorithm found at least a feasible solution (that is, an assignment of values to variables that satisfies the constraints of the model, though it may not necessarily be optimal). Unknown specifies that the algorithm has no information about the solution of the model. Optimal specifies that the algorithm found an optimal solution (that is, an assignment of values to variables that satisfies all the constraints of the model and that is proved optimal with respect to the objective of the model).

\section{Cognitive Complexity Reducer Approach}
\label{sec:approach}
We proposed a SSCC reducer approach consisting in a solver that takes as input the path to a software project and the cognitive complexity threshold ($\tau$) \cite{Saborido2022}. For each method with SSCC greater than $\tau$, it searches for sequences of applicable Extract Method refactoring operations. Finally, it outputs the changes to perform to each method. In order to search for Extract Method refactoring opportunities in a method, this approach generates its corresponding Abstract Syntax Tree (AST). Second, it parses the AST and annotates different properties in each node: its contribution to the SSCC of the method, the accumulated value of the inherent component ($\iota$), the accumulated value of the nesting component ($\nu$), the number of elements contributing to the nesting component of the SSCC of the node ($\mu$), and its nesting component ($\lambda$). Third, the approach processes the annotated AST to compute the list of consecutive sentences contributing to the SSCC of the method. This is done to obtain Extract Method refactoring opportunities. Although sentences contributing to the SSCC must be part of code extractions, it is also necessary to consider single statements even if they do not contribute to the value of this metric. The inclusion of statements of this kind could suppose that the extraction is feasible or not. For example when several arithmetic operations are needed to compute a result, if all operations are not included in the extraction, the refactoring probably is not possible because only one variable could be returned.  
Once the approach identifies Extract Method refactoring opportunities, it checks if the extractions are applicable. This is done with the help of refactoring tools which are able to check pre-conditions, post-conditions, and apply the corresponding operation over the source code.  

\subsection*{Cognitive Complexity Reducer Tool Implementation}
We proposed a Java cognitive complexity reducer tool as an Eclipse application \cite{Saborido2022}. The goal is to provide the necessary means for generating an Eclipse product that can be run from the operating system command-line as a standalone executable, without the need for opening Eclipse for running. This is particularly useful if, for instance, one needs to integrate it in their current development workflow (e.g., using continuous integration). We got this idea from the jDeodorant project\footnote{\url{https://github.com/tsantalis/JDeodorant}}, an Eclipse plug-in that detects design problems in Java software and recommends appropriate refactorings to resolve them.

\begin{figure}[!htb]
    \centering
    \includegraphics[scale=0.37]{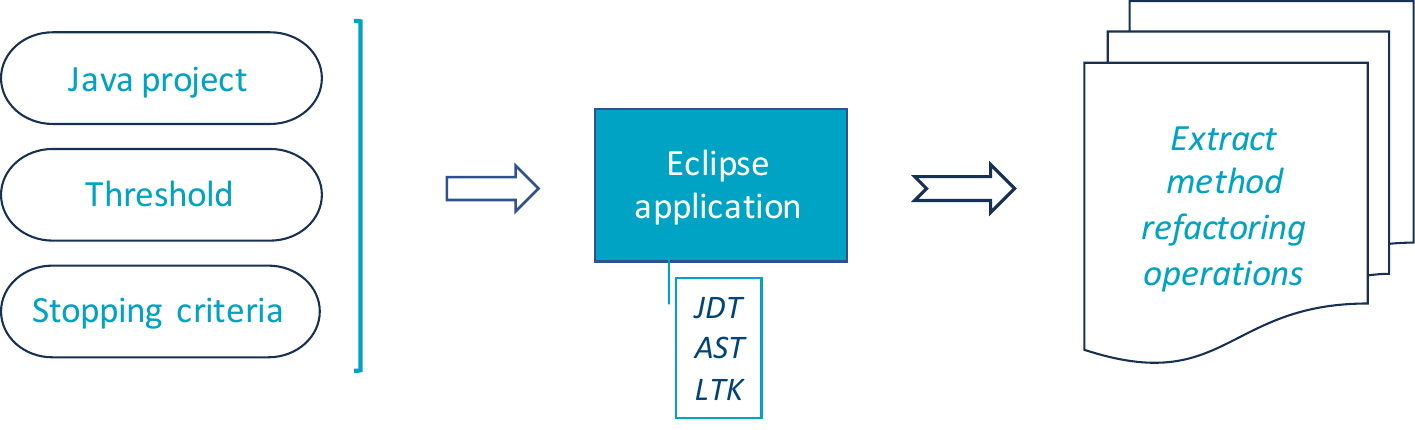}
    \caption{Overview of the proposed cognitive complexity reducer approach. JDT and LTK stand for the Eclipse Java Development Toolkit and Language Toolkit, respectively.}
    \label{fig:eclipse-app}
\end{figure}

The developed tool takes as input (i) a SonarQube server URL, (ii) the path to the software project to process, (iii) the cognitive complexity threshold ($\tau$), and (iv) the stopping criterion (number of seconds). Then, it gets all existing cognitive complexity issues by querying the SonarQube web API (assuming that an analysis of the project already exists in the server). Next, for each method with SSCC greater than $\tau$, it computes the refactoring cache, generates the conflict graph, and searches for optimal Extract Method refactoring opportunities. Fig.~\ref{fig:eclipse-app} summarizes the proposed approach.

From the conflict graph of a method, the tool models the ILP problem as follows: 
\begin{itemize}
    \item Each node in the conflict graph defines a binary variable $x_i,  \forall i = 0,1, \ldots, m$ (Eq.~\ref{eqn:domain-x}).
    \item Each conflict edge introduces a new constraint (Eq.~\ref{eqn:conflict}).
    \item Each node $i, \; \forall i = 0,1,\ldots,m$, in the conflict graph, introduces new constraints involving nested nodes (i.e., nodes $j \; | \; j \rightarrow i, \; \forall i = 0,1,\ldots,m$) (Eq.~\ref{eqn:limit}) and intermediate nodes (i.e., nodes $l \; | \; j \rightarrow l \rightarrow i, \; \forall j = 0,1,\ldots,m$) (Eq.~\ref{eqn:zdef}).
\end{itemize}

The resulting model is then solved using CPLEX as ILP solver, in order to compute the optimal solution(s). %which involves assigning a value of 0 or 1 to the variables $X_i$ that represent the nodes to be extracted in order to reduce cognitive complexity.
The resulting ILP model associated to the illustrating example has 50 decision variables and 54 constraints. It is shown in \mbox{Listing \ref{listingILPModelIllustratingExample}} of the Appendix.

\section{Case Study}
\label{sec:casestudy}
In this section, we describe the study we conduct to evaluate the proposed approach when reducing the SSCC of 10 open-source projects. Next, we detail the objects of study. Then, we describe the experimental setup.

\subsection{Objects of Study}
We use a diverse set of 10 open-source projects selected from our previous work \cite{Saborido2022}: two popular frameworks for multi-objective optimization, five platform components to accelerate the development of smart solutions, and three popular open-source projects with more than 10,000 stars and forked more than 900 times. \mbox{Table~\ref{tab:projects-metrics}} shows these projects and some software metric values. In order to ease the replication of the study, for each software project we also show its abbreviated commit hash on GitHub.

\begin{table}[!htb]
\centering
\scriptsize
\caption{Case study projects metrics: name of the open-source project (its abbreviated commit hash), number of classes, number of methods, lines of code (LOC), and number of cognitive complexity issues reported by SonarQube, respectively.}
\label{tab:projects-metrics}
\begin{tabular}{lrrrr}
\hline
Project {\tiny (Commit)}       & \#Classes & \#Methods & LOC     & \#CC issues \\ \hline
ByteCode {\tiny (55bfc32)}   & 301       & 1,350     & 23,071  & 57 (4\%)       \\
CyberCaptor {\tiny (b6b1f10)}  & 85        & 784       & 17,023  & 37 (5\%)      \\
FastJson {\tiny (93d8c01e9)}      & 256       & 2,039     & 43,644  & 230  (11\%)    \\
Fiware-Commons {\tiny (f83b342)} & 28        & 155       & 1,325   & 4   (3\%)     \\
IoTBroker   {\tiny (98eeceb)}   & 79        & 646       & 7,940   & 10  (2\%)     \\
Jedis  {\tiny (cfc227f7)}        & 295       & 1,917     & 16,566  & 3  ($<$1\%)      \\
jMetal  {\tiny (e6baf75aa)}       & 610       & 3,327     & 43,298  & 63 (2\%)      \\
Knowage-core {\tiny (dfed28a869)}  & 1,093     & 6,967     & 149,137 & 558 (8\%)     \\
MOEA-framework {\tiny (223393fd)} & 506       & 2,939     & 33,888  & 82  (3\%)     \\
QueryExecution {\tiny (c032e5a)} & 6         & 53        & 1,013   & 6    (11\%)    \\ 
\hline
Total & 3,259 & 20,177 & 336,905 & 1,050 \\
\hline
\hline
\end{tabular}
\end{table}

The simplest open-source project is QueryExecution: it contains 53 methods and six classes, summing up 1,013 lines of code. Despite the low number of methods in comparison to other open-source projects, 6 over 53 (11\%) of the methods of QueryExecution have SSCC greater than 15 (the default threshold). Although this project looks simple, and, therefore, easy to maintain, reducing the SSCC of these six methods is not straight forward. For instance, for the method \texttt{getDBIds} SonarQube suggests reducing its SSCC from 41 to 15. However, there are several refactoring opportunities that can be applied to get this done. Conversely, Knowage-core is the most complex project in our case study: it contains 6,967 methods and 1,093 classes, summing up 149,137 lines of code. Even for a senior developer, maintaining this ecosystem is complicated and prone to errors. SonarQube reports 558 cognitive complexity issues for this project, i.e., 8\% of the methods in the project have SSCC greater than 15. Reducing the SSCC of these 558 methods would be time-consuming and prone to errors when done manually.

We validate the proposed cognitive complexity reduction tool over the 10 open-source projects shown in \mbox{Table \ref{tab:projects-metrics}}. In total, these projects have 1,050 cognitive complexity issues. The goal of the study is to validate to which extent the proposed approach is able to reduce the number of cognitive complexity issues existing on these projects. In addition, we want to uncover how many extractions are needed, how many lines of codes are extracted, and how many parameters new extracted methods have when reducing the SSCC of methods.

\subsection{Experimental Setup}
We conducted the experiments in a laptop MacBook Air with Apple M1 chip and 8 GiB of RAM, running the operating system macOS Ventura. We used SonarQube version 7.2 and the Eclipse IDE version 2022-09 (4.25.0). We set the cognitive complexity threshold to the default value proposed by SonarQube ($\tau=15$). AST processing and Extract Method refactorings were performed through Eclipse JDT version 3.31.0 and Eclipse LTK version 3.10.0. All graph generation in our tool has been developed using the jGraphT library, a Java library of graph theory data structures and algorithms.\footnote{\url{https://jgrapht.org/}}
In order to solve ILP optimization problems, we used the java library of CPLEX provided by the IBM ILOG CPLEX Optimization Studio V22.1.0 through the IBM Academic Initiative.

\section{Results}
\label{sec:results}
Table~\ref{tab:modelsStats} summarizes ILP model stats for the 1,050 methods under study. On average, the number of variables and constraints varies from 125 to 11,208 and from 129 to 16,253, respectively. However, the ILP model of some methods contains thousands or even millions of variables and/or constraints.

Table~\ref{tab:modelStatus300seconds} shows CPLEX model status when using 300 seconds as stopping criterion for the 1,050 methods under study. Aggregating all projects, CPLEX solver found optimal solutions in 885 (84\%) of the 1,050 methods. For the rest of methods, 165 (16\%), CPLEX did not find any optimal solution. Over these 165 methods, (i) CPLEX reported solutions as non-optimal (feasible model) for 23 (14\%) methods after meeting the stopping criterion, (ii) CPLEX reported no solution (unknown model) for 15 (9\%) methods after meeting the stopping criterion, (iii) CPLEX reported no solution (infeasible model) for 109 (66\%) methods before meeting the stopping criterion, and (iv) CPLEX threw an \texttt{OutOfMemoryException} exception when creating the model or solving the problem for 18 (11\%) methods.

\begin{table}[!htb]
\centering
\caption{CPLEX model status when  using 300 seconds as stopping criterion for the 1,050 methods under study. The last column refers to methods for which CPLEX threw a Java \texttt{OutOfMemoryException} exception when creating the model or solving the problem.}
\begin{tabular}{
m{0.13\textwidth}
>{\raggedleft}m{0.04\textwidth}
>{\raggedleft}m{0.04\textwidth}
>{\raggedleft}m{0.05\textwidth}
>{\raggedleft}m{0.05\textwidth}
r
}
\hline
Project        & Optimal    & Feasible & Unknown  & Infeasible & Error    \\ 
\hline
ByteCode       & 52 %(91.23\%)
& 0 %(0.00\%) 
& 0 %(0.00\%) 
& 4 %(7.02\%)   
& 1 %(1.75\%)  
\\ 
CyberCaptor    & 33 %(89.19\%) 
& 0 %(0.00\%)  
& 0 %(0.00\%) 
& 4 %(10.81\%)  
& 0 %(0.00\%)  
\\
FastJson       & 147 %(63.91\%)
& 4 %(1.74\%)  
& 6 %(2.61\%)  
& 72 %(31.30\%) 
& 1 %(0.43\%) 
\\
Fiware-Commons & 3 %(75.00\%) 
& 0 %(0.00\%)  
& 0 %(0.00\%)
& 1 %(25.00\%)  
& 0 %(0.00\%)
\\
IoTBroker      & 10 %(100.00\%) 
& 0 %(0.00\%)  
& 0 %(0.00\%)  
& 0 %(0.00\%)    
& 0 %(0.00\%) 
\\
Jedis          & 3 %(100.00\%)
& 0 %(0.00\%)  
& 0 %(0.00\%) 
& 0 %(0.00\%)  
& 0 %(0.00\%)  
\\
jMetal         & 59 %(93.65\%)   
& 3 %(4.76\%) 
& 0 %(0.00\%) 
& 1 %(1.59\%)  
& 0 %(0.00\%) 
\\
Knowage-core   & 494 %(88.53\%)
& 15 %(2.69\%)
& 9 %(1.61\%)
& 24 %(4.30\%) 
& 16 %(2.87\%) 
\\
MOEA-framework          & 78 %(95.12\%)  
& 1 %(1.22\%)
& 0 %(0.00\%)  
& 3 %(3.66\%)  
& 0 %(0.00\%) 
\\
QueryExecution & 6 %(100.00\%) 
& 0 %(0.00\%)  
& 0 %(0.00\%) 
& 0 %(0.00\%) 
& 0 %(0.00\%)  
\\
\hline
Total          & 885 & 23  & 15  & 109 & 18  \\
(Percentage) & 84\% & 2\% & 1\% & 10\% & 2\% \\
\hline
\hline
\end{tabular}
\label{tab:modelStatus300seconds}
\end{table}

We performed further experiments setting CPLEX stopping criterion to 3,600 seconds (one hour) for those 38 methods (23 + 15) for which CPLEX reported the model as feasible or unknown after 300 seconds.  Table~\ref{tab:modelStatus3600seconds} summarizes the results. Over these 38 methods, (i) CPLEX found optimal solutions for nine (24\%) methods, (ii) CPLEX reported solutions as non-optimal (feasible model) for 14 (37\%) methods, (iii) CPLEX reported no solution (unknown model) for 12 (32\%) methods, and (iv) CPLEX reported no solution (infeasible model) for three (8\%) methods.

\begin{table}[!htb]
\centering
\caption{CPLEX model status when using 3,600 seconds as stopping criterion for the 38 methods (23 + 15) which CPLEX reported the model as feasible or unknown after 300 seconds.}
\begin{tabular}{llll}
\hline
Optimal  & Feasible  & Unknown  & Infeasible \\ \hline
9 (24\%) & 14 (37\%) & 12 (32\%) & 3 (8\%) \\ 
\hline
\hline
\end{tabular}
\label{tab:modelStatus3600seconds}
\end{table}

We finally studied those methods for which CPLEX returned the \texttt{OutOfMemoryException} exception when creating the ILP model or solving the ILP problem. Table~\ref{tab:methodsOutOfMemory} shows the name of the project and the Java class each method belongs to, the cognitive complexity of the method, the LOC, and the number of variables and constraints of their models. As shown, these methods ILP problems contain more than 50,000 variables and/or more than 89,000 constraints.

\begin{table*}[!htb]
\centering
\caption{CPLEX model stats for methods under study per project.}
\begin{tabular}{l|rrrrrr|rrrrrr}
\hline
\multirow{2}{*}{project} & \multicolumn{6}{c}{\#Variables}                     & \multicolumn{6}{c}{\#Constraints}                   \\
                         & Min. & 1st.Q. & Median & Mean  & 3rd.Q. & Max     & Min. & 1st.Q. & Median & Mean  & 3rd.Q. & Max     \\
\hline
ByteCode                 & 5    & 153    & 295    & 1,503  & 831    & 15,749   & 6    & 159    & 304    & 1,983  & 915    & 27,393   \\
CyberCaptor              & 1    & 62     & 352    & 3,048  & 1,375   & 52,993   & 2    & 64     & 417    & 4,669  & 1,962   & 96,665   \\
FastJson                 & 1    & 65     & 167    & 8,213  & 497    & 1,164,599 & 2    & 70     & 171    & 13,149 & 521    & 2,151,057 \\
Fiware-Commons           & 15   & 20     & 124    & 133   & 238    & 270     & 16   & 21     & 127    & 137   & 243    & 277     \\
IoTBroker                & 15   & 56     & 120    & 125   & 179    & 295     & 16   & 58     & 126    & 129   & 183    & 298     \\
Jedis                    & 88   & 98     & 108    & 290   & 391    & 673     & 92   & 101    & 109    & 297   & 400    & 690     \\
jMetal                   & 5    & 212    & 600    & 2,825  & 1,946   & 31,961   & 6    & 236    & 717    & 3,898  & 2,334   & 61,515   \\
Knowage-core             & 1    & 330    & 1007   & 11,208 & 4,391   & 672,131  & 2    & 352    & 1,147   & 16,253 & 5,332   & 978,567  \\
MOEAFramework            & 10   & 169    & 350    & 2,540  & 1,160   & 63,311   & 11   & 182    & 371    & 3,938  & 1,498   & 117,836  \\
QueryExecution           & 170  & 299    & 427    & 749   & 504    & 2,685    & 172  & 317    & 484    & 836   & 536    & 3,062   
\\
\hline
\hline
\end{tabular}
\label{tab:modelsStats}
\end{table*}

\begin{table*}[!htb]
\caption{Methods for which CPLEX returned the \texttt{OutOfMemoryException} exception when creating the ILP model or solving the ILP problem. It shows the initial cognitive complexity of the method, its LOC, and the number of variables and constraints of the corresponding ILP problem. The '-' symbol is used for those methods for which CPLEX was unable to create the ILP model.}
\centering
\begin{tabular}{lllrrrr}
\hline
Project      & Class                              & Method               & initialCC & LOC & \#Variables & \#Constraints \\
\hline
ByteCode     & \texttt{MainViewerGUI.java}                 & \texttt{buildSettingsMenu}    & 19        & 213 & -         & -           \\
FastJson     & \texttt{ASMDeserializerFactory.java}        & \texttt{\_deserialze}         & 80        & 457 & 1,164,599 & 2,151,057   \\
Knowage-core & \texttt{BarCharts.java}                     & \texttt{configureChart}       & 145       & 332 & 209,987   & 348,432     \\
Knowage-core & \texttt{CombinedCategoryBar.java}           & \texttt{createChart}          & 109       & 291 & 426,982   & 670,592     \\
Knowage-core & \texttt{DataSetJSONSerializer.java}         & \texttt{serialize}            & 129       & 309 & 672,131   & 978,567     \\
Knowage-core & \texttt{DatasetWizardTag.java}              & \texttt{doStartTag}           & 22        & 262 & -         & -           \\
Knowage-core & \texttt{ListTag.java}                       & \texttt{makeNavigationButton} & 71        & 240 & 410,697   & 525,289     \\
Knowage-core & \texttt{ListTag.java}                       & \texttt{makeRows}             & 227       & 279 & 214,601   & 271,503     \\
Knowage-core & \texttt{LovWizardTag.java}                  & \texttt{doStartTag}           & 24        & 247 & -         & -           \\
Knowage-core & \texttt{MenuListJSONSerializerForREST.java} & \texttt{createEndUserMenu}    & 25        & 266 & 325,750   & 621,888     \\
Knowage-core & \texttt{OverlaidBarLine.java}               & \texttt{createChart}          & 140       & 351 & 186,068   & 230,648     \\
Knowage-core & \texttt{OverlaidStackedBarLine.java}        & \texttt{createChart}          & 59        & 147 & 64,661    & 89,395      \\
Knowage-core & \texttt{SelectParametersLookupModule.java}  & \texttt{loadList}             & 63        & 229 & 269,673   & 359,626     \\
Knowage-core & \texttt{SimpleBar.java}                     & \texttt{createChart}          & 34        & 114 & 65,715    & 106,665     \\
Knowage-core & \texttt{SparkLine.java}                     & \texttt{createChart}          & 51        & 146 & 130,439   & 218,092   \\
Knowage-core & \texttt{SpeedometerMultiValue.java}         & \texttt{createChart}          & 33        & 90  & 50,640    & 91,836      \\
Knowage-core & \texttt{StackedBar.java}                    & \texttt{createChart}          & 52        & 179 & -         & -           \\
Knowage-core & \texttt{StackedBarGroup.java}               & \texttt{createChart}          & 49        & 137 & 136,893   & 227,233     \\
\hline
\hline
\end{tabular}
\label{tab:methodsOutOfMemory}
\end{table*}

\subsection{Running Times and Perfomance}
Using 300 seconds as stopping criterion, the proposed approach is able to find optimal solutions for 885 (84\%) cognitive complexity issues. Over the 1,050 methods under study, CPLEX ended before meeting the stopping criterion for 994 (95\%) methods (excluding those 18 methods for which CPLEX threw the runtime exception). This means that CPLEX is able to find optimal solutions or to determine the infeasibility of the model in less than 300 seconds for most methods. Increasing the stopping criterion by a factor of 12 (3,600 seconds), the proposed approach is able to fix nine cognitive complexity issues more. Based on the previous, setting the stopping criterion to 300 seconds is enough for 95\% of the methods under study.

Table \ref{tab:executionTimeStats300seconds} shows execution time stats for computing optimal solutions when using 300 seconds as stopping criterion. CPLEX finds optimal solutions in less than five seconds, on average. Although our approach took almost two hours to fill the refactoring cache in the worst case, it usually performs this task in 39 seconds, on average. Regarding the 3rd quartile, filling the refactoring cache took less than 22 seconds for the 75\% of the 885 methods for which CPLEX found optimal solutions.

\begin{table}[!htb]
\caption{Execution times stats (in seconds) for computing optimal solutions when using 300 seconds as stopping criterion.}
\label{tab:executionTimeStats300seconds}
\centering
\begin{tabular}{l|rr}
\hline
Metric        & CPLEX runtime & Time to fill cache  \\
\hline
Min.    & 0.002  & 0.292 \\
1st.Q. & 0.042  & 4.844 \\
Median  & 0.120  & 9.719 \\
Mean    & 4.047  & 38.974 \\
3rd.Q. & 0.538  & 21.807 \\
Max.    & 372.562 & 7,135.624 \\
\hline
\hline
\end{tabular}
\end{table}

\subsection{Extract Method Refactoring Operations Metrics}
\begin{figure*}[!htb]
    \includegraphics[scale=0.45]{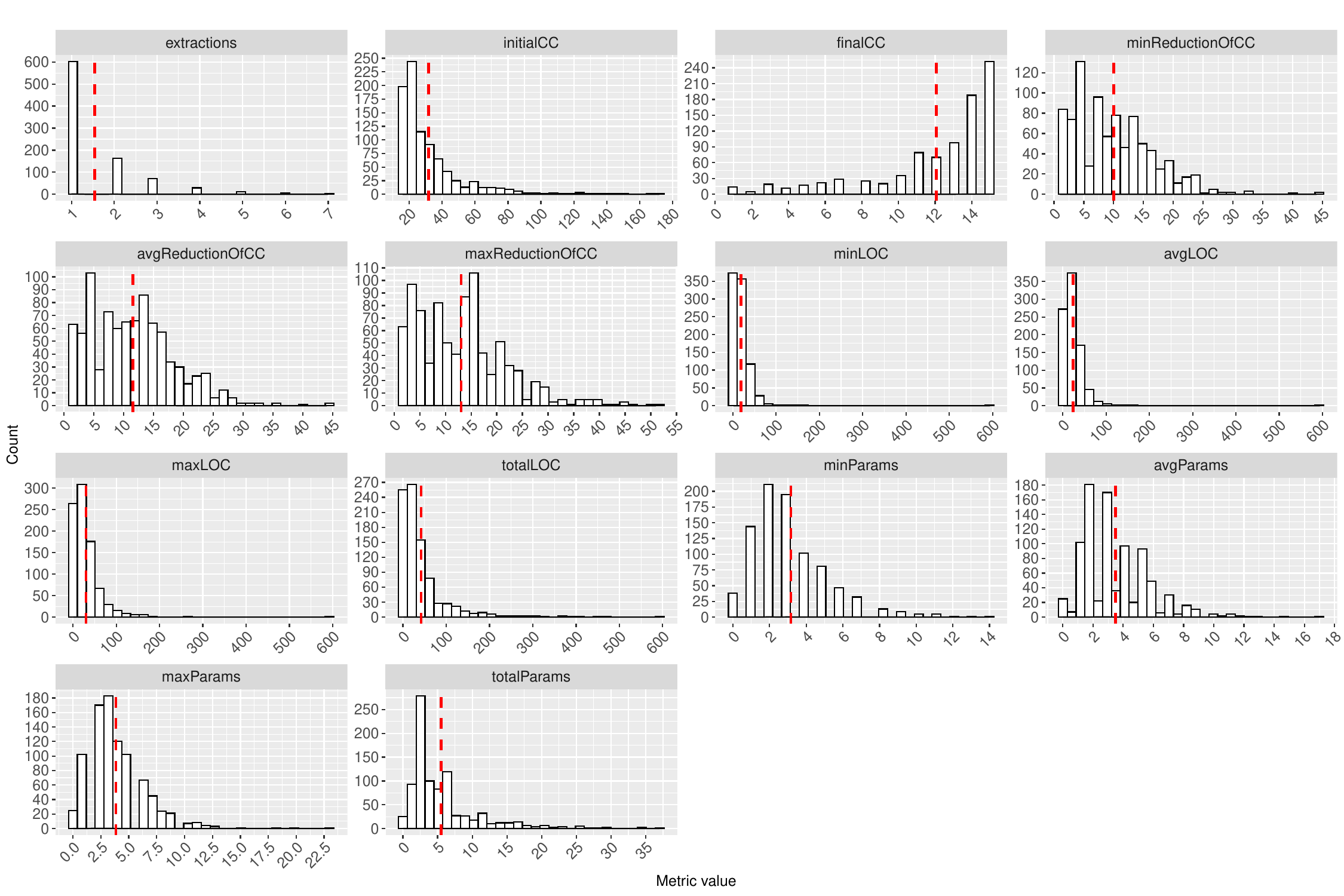}
    \caption{Histogram of studied metrics for the 885 methods for which CPLEX found optimal solutions when using 300 seconds as stopping criterion. Vertical dashed red lines correspond to the mean value of each metric.}
    \label{fig:Histogram}
\end{figure*}

Fig.~\ref{fig:Histogram} and Table~\ref{tab:metricValueStats300seconds} show the frequency distribution and  metric value stats of the 885 methods for which CPLEX found optimal solutions when using 300 seconds as stopping criterion, respectively. \texttt{extractions} refers to the number of code extractions required to reduce cognitive complexity. \texttt{initialCC} and \texttt{finalCC} are the initial and final cognitive complexity of methods, respectively. \texttt{minCCR}, \texttt{avgCCR}, and \texttt{maxCCR} refer to the minimum, average, and maximum cognitive complexity reduction associated to code extraction(s) of a method, respectively. \texttt{minLOC}, \texttt{avgLOC}, and \texttt{maxLOC} correspond to the minimum, average, and maximum number of extracted lines of code (LOC) associated to code extraction(s) of a method, respectively. \texttt{totalLOC} refers to the cumulative number of extracted lines of code of all code extractions in a method. \texttt{minParams}, \texttt{avgParams}, and \texttt{maxParams} are the minimum, average, and maximum number of parameters of the extracted methods  when reducing the cognitive complexity of a method, respectively. \texttt{totalParams} refers to the cumulative number of parameters of the extracted methods. As shown, one extraction is enough to reduce the cognitive complexity of 602 (68\%) methods. However, 283 (32\%) methods require more than one extraction to reduce their cognitive complexity. We also observe that no more than two extractions are required for 75\% of the methods (3rd.Q column). The cognitive complexity reduction varies from one to 51, averaging a reduction of 11 by code extraction. The number of extracted LOC varies from one to 594, averaging 23 LOC by extraction. Regarding the number of parameters of new methods, this varies from zero to 23, averaging three parameters.

\begin{table}[!htb]
\caption{Metric value stats of optimal solutions found by CPLEX when using 300 seconds as stopping criterion.}
\label{tab:metricValueStats300seconds}
\begin{tabular}{l|rrrrrr}
\hline
Metric                      & Min.    & 1st.Q. & Median  & Mean    & 3rd.Q. & Max.     \\
\hline
extractions                   & 1  & 1.00  & 1  & 1.54  & 2.00  & 7  \\
\hline
initialCC             & 16 & 19.00 & 25 & 31.66 & 36.00 & 172 \\
finalCC               & 1  & 11.00 & 13 & 12.06 & 15.00 & 15  \\
minCCR              & 1  & 5.00  & 9  & 10.02 & 14.00 & 44  \\
avgCCR             & 1  & 5.00  & 11 & 11.52 & 15.50 & 44  \\
maxCCR              & 1  & 6.00  & 12 & 13.08 & 18.00 & 51 \\
\hline
minLOC               & 1  & 6.00  & 13 & 18.76 & 24.00 & 594 \\
avgLOC              & 1  & 8.50  & 19 & 23.80 & 31.67 & 594 \\
maxLOC               & 1  & 9.00  & 21 & 30.04 & 39.00 & 594 \\
totalLOC             & 1  & 9.00  & 24 & 42.32 & 48.00 & 594 \\
\hline
minParams     & 0  & 2.00  & 3  & 3.16  & 4.00  & 14  \\
avgParams    & 0  & 2.00  & 3  & 3.49  & 5.00  & 17  \\
maxParams     & 0  & 2.00  & 3  & 3.84  & 5.00  & 23  \\
totalParams   & 0  & 2.00  & 4  & 5.50  & 7.00  & 37  \\
\hline
\hline
\end{tabular}
\end{table}

\begin{figure*}[!htb]
    \includegraphics[scale=0.48]{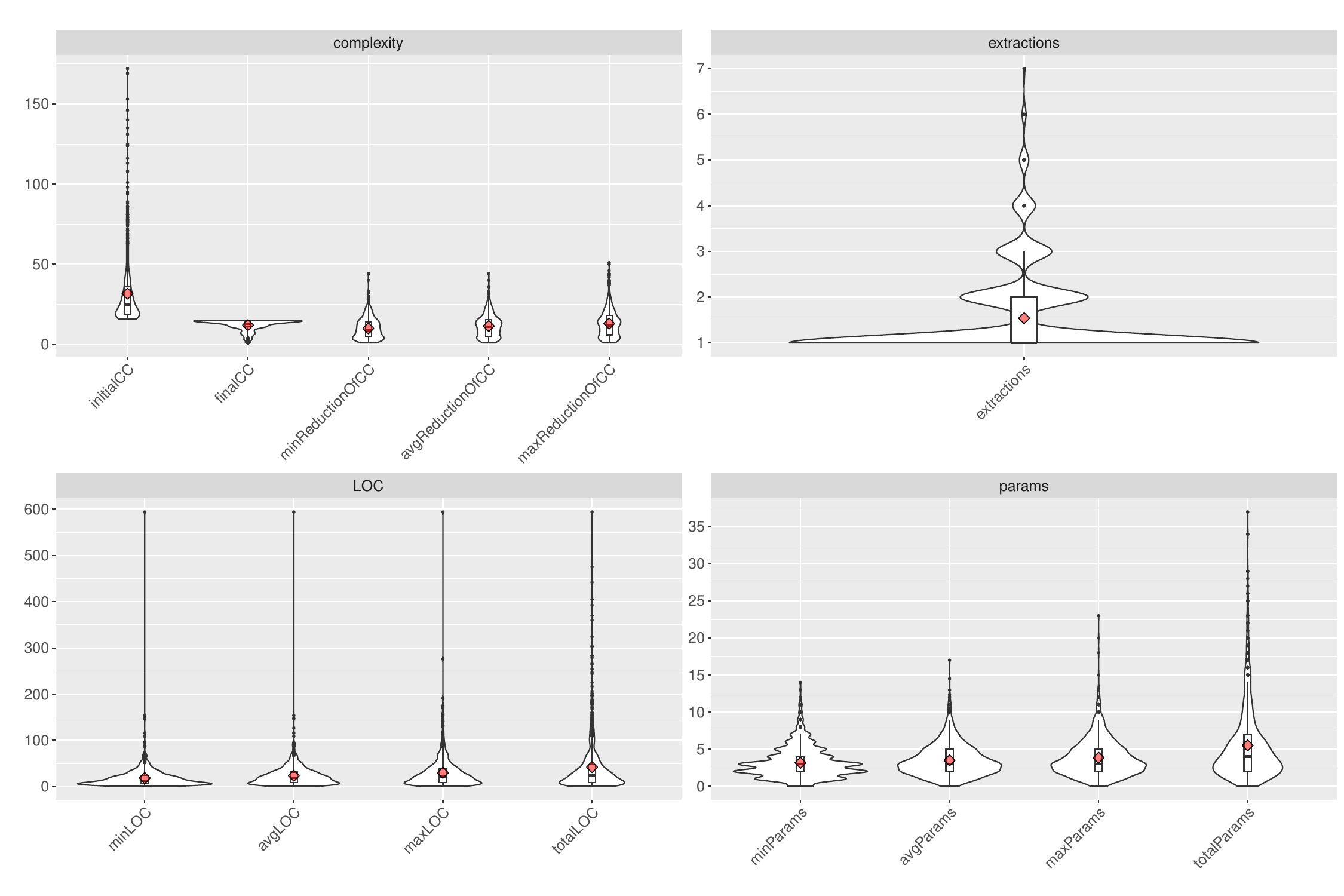}
    \caption{Violin plots of studied metrics for the 885 methods for which CPLEX found optimal solutions when using 300 seconds as stopping criterion. Red diamonds correspond to the mean value of each metric.}
    \label{fig:ViolinplotByMetricsCategory}
\end{figure*}

Fig.~\ref{fig:ViolinplotByMetricsCategory} shows summary statistics and the density of studied metrics though violin plots, grouping related metrics in different categories: \texttt{complexity}, \texttt{extractions}, \texttt{LOC}, and \texttt{params}. Violin plots are similar to box plots, except that they also show the probability density of the data at different values, usually smoothed by a kernel density estimator. Thus, this plot complements Fig.~\ref{fig:Histogram} providing additional information about the probability density of each metric at all observed values in the study. 
For completeness, we show frequency distribution of metrics and summary statistics for each individual software project under study in our appendix (Figures \ref{fig:Histogram-ByteCode}-\ref{fig:Violinplot}).

\begin{figure}[!htb]    
\includegraphics[scale=0.55]{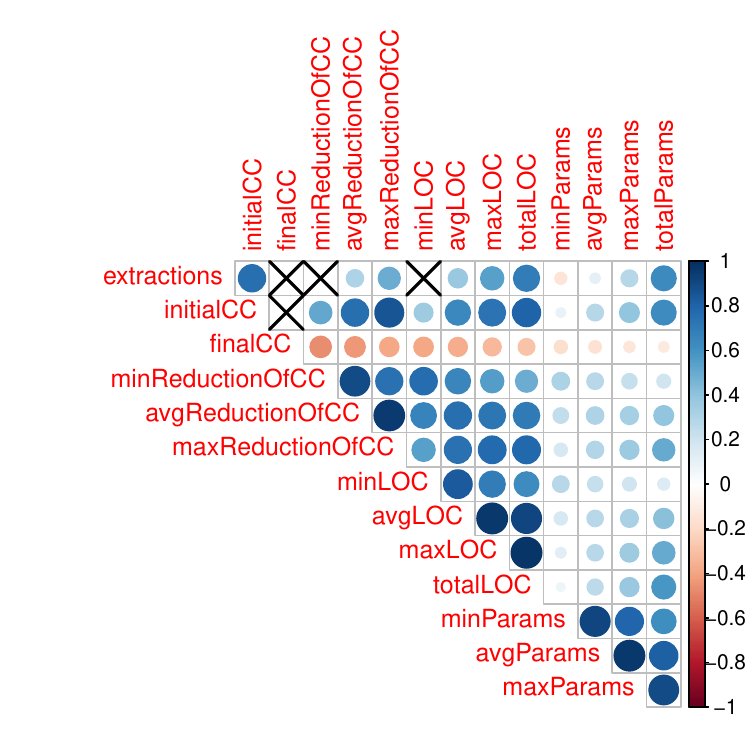}
\caption{Correlation between metrics under study.}
\label{fig:correlation}
\end{figure}

We finally analyze the correlation between metrics under study. Fig. \ref{fig:correlation} shows the Spearman correlation, using black crosses on non-significant coefficients (using 0.05 as significant level). As expected, higher the initial cognitive complexity, higher the number of extractions, the cognitive complexity reduction, and the number of extracted LOC.  

\section{Discussion}
\label{sec:discussion}
We were able to reduce the cognitive complexity to or below the default threshold (15) in 78\% of the methods under study by using exhaustive search  \cite{Saborido2022}. On average, enumerations algorithms took less than 70 seconds per method for finding feasible solutions to the cognitive complexity reduction problem. However, enumerating sequences of code extractions cannot assure solutions optimality nor that no solution exist. The reason is that the cost of exploring all possible sequences of extractions might be unaffordable in some cases. By modeling the cognitive complexity reduction task as an ILP optimization problem, CPLEX was able to find optimal solutions or to determine the unfeasibility of ILP models in less than five seconds, on average. Using 300 seconds as stopping criterion, the proposed approach found optimal solutions for most cognitive complexity issues: 885 (84\%) of those 1,050 methods under study. Thus, setting the stopping criterion to 300 seconds was enough for 95\% of the methods under study.

\section{Threats to validity}
\label{sec:threats}
This section discusses all  threats that might have an impact on the validity of our study following common guidelines for empirical studies~\cite{Yin2002}.

Threats to internal validity concern factors that could have influenced our results. A possible threat to internal validity is the stopping criterion set to 300 seconds for CPLEX. This stop condition might have influenced our results, because larger times could be ending up solving more cognitive complexity issues. In order to alleviate this threat, we performed additional experiments setting the stopping criterion to 3,600 seconds in Section~\ref{sec:casestudy}. However, we found that setting the stopping criterion to 300 seconds is enough for 95\% of the methods under study. Another aspect that can influence the results is the choice of the cognitive complexity threshold used. We used the cognitive complexity threshold suggested by the static code tools SonarCloud and SonarQube.

Threats to construct validity concern the relationship between theory and observation and the extent to which the measures represent real values. In our study, all the experiments were run in the same computer and the metrics we collected are all consistent when analyzing  the original and the resulting source codes.

Threats to external validity concern the generalization of our findings. To reduce external validity threats, we selected a diverse set of 10 open-source projects for our case of study. Aggregating all projects, we processed 1,050 methods with SSCC greater than 15. This high number of existing issues guarantees that we have analyzed very diverse methods in terms of complexity and size. Thus, we guess our findings can be generalized to other software projects. 

Threats to conclusion validity concern the relationship between experimentation and outcome. Many methods were analyzed and optimal solution were found for most of the methods of the software projects under study.

\section{Related work}
\label{sec:relatedWork}
Probably the oldest and most intuitively obvious notion of software complexity is the number of statements in the program, or the statement count. However, numerous software complexity measures have been proposed in the past: the number of program statements, McCabe’s cyclomatic number \cite{McCabe1976}, Halstead’s programming effort \cite{Halstead1977}, and the Knot measure \cite{Woodward1979} were the most frequently cited measures in the 70s. In the 90s, Douce \etal introduced a set of metrics
that help in calculating the complexity of a given system or
program code based on the object-oriented concepts such as
the object and class \cite{DBLP:conf/ppig/DouceLB99}. All those metrics were based on the spatial abilities, which measure the complexity by calculating the distances between the program elements in the code.

Misra proposed an object-oriented complexity metric which calculates the complexity of a class at method level \cite{Misra2007b}, in 2007. Later, in 2008, Misra \etal proposed a metric that considers internal attributes which directly affect the complexity of software: number of lines, total occurrence of
operators and operands, number of control structures,
and function calls (coupling) \cite{Misra2008}. The same year, Misra and Akman proposed a new complexity metric based on cognitive informatics for object-oriented code covering cognitive complexity of the system, method complexity, and complexity due to inheritance together \cite{Misra2008b}.

Few years later, in 2011, Misra \etal proposed a cognitive complexity metric for evaluating design of object-oriented code. The proposed metric is based on the inheritance feature of the object-oriented systems. It calculates the complexity at method level considering internal structure of methods, and also considers inheritance to calculate the complexity of class hierarchies \cite{MISRA2011}. In 2012, Misra \etal proposed a suite of cognitive metrics for evaluating complexity of object-oriented codes \cite{misra_suite_2012}. 
The same year, Misra \etal also proposed a framework for the evaluation and validation of software complexity measure. This framework is designed to analyze whether software metric qualifies as a measure from different perspectives \cite{iet-sen.2011.0206}.

In 2016, Haas and Hummel addressed the problem of finding the most appropriate refactoring candidate for long methods written in Java. The approach determines valid refactoring candidates and ranks them using a scoring function that aims to improve readability and reduce code complexity \cite{Haas2016}. 
Later that year, Wijendra and Hewagamage proposed a cognitive complexity metric which determines the amount of information inside the software through cognitive weights and the way of information scattering in terms of Lines of Code (LOC) \cite{wijendra_automated_2016}. The same year, Crasso \etal presented a software metric to assess cognitive complexity in object-oriented systems developed in the Java language \cite{crasso_assessing_2016}. The proposed metric is based on a characterization of basic control structures present in Java systems. Authors also provided several algorithms to compute the metric and introduced their materialization in the Eclipse IDE.

In 2018, SonarSource introduced cognitive complexity as a new metric for measuring the understandability of any given piece of code \cite{Campbell2018}. This paper investigated developers' reaction to the introduction of cognitive complexity in the static code analysis tool service SonarCloud. In an analysis of 22 open-source projects, they assessed whether a development team `accepted' the proposed metric based on whether they fixed code areas of high cognitive complexity as reported by the tool. They found that the metric had a 77\% acceptance rate among developers.

Kaur and Mishra conducted an experimental analysis in which the software developer's level of difficulty in comprehending the software was theoretically computed and empirically evaluated for estimating its relevance to actual software change \cite{kaur_cognitive_2019}, in 2019. 
This study validated a cognitive complexity metric as a noteworthy measure of version to version source code change. Also in 2019, Alqadi proposed novel metrics to compute the cognitive complexity of code slices \cite{alqadi_relationship_2019}. 
Empirical investigation into how cognitive complexity correlates with defects in the version histories of three open-source systems was performed. The results showed that the increase of cognitive complexity significantly increases the number of defects in 93\% of the cases. The same year, Hubert proposed an approach to fully automate the extract method refactoring task, ranking refactoring opportunities according to a scoring function which takes into account software cognitive complexity \cite{Hubert2019}.

In 2020, Muñoz Barón \etal conducted a systematic literature search to obtain data sets from studies which measured code understandability and found that cognitive complexity integrated in the well-known static code analysis tool service SonarCloud positively correlates with comprehension time and subjective ratings of understandability \cite{munoz_baron_empirical_2020}.

Recently, in 2022, Wijendra \etal, proposed to convert the source code logical behavior into its graphical representation with the usage of control flow graphs \cite{Wijendra2022}. This allows users to navigate through the graphical representation and comprehend the logic of the source code. It has been observed, through four different user groups with different academic levels, that the time taken to maintain software with the graphical representation is lesser than its code base referring. This implies the comprehension effort and the cognitive complexity reduction can be achieved with the aid of diagrammatic views. Wijendra \etal have also evaluated the impact of refactoring techniques on software understandability \cite{Wijendra2022b}. The non-refactored and refactored source codes of a software project were given to a sample user group of 500 programmers to modify the software according to a set of requirements by understanding the logic behind it. The time taken by each user was measured for both refactored and non-refactored source code modifications. According to the time taken to modify the source code in five different situations, authors concluded that less time has been spent to modify the refactored source code rather than implementing the same in non-refactored code in most of the cases. However, when the Extract Method refactoring technique was applied, users required longer times because they had to select the set of lines of codes to be extracted as a separate method. Also in 2022, Akalanka \etal proposed a system to automate certain functionalities in the source code to reduce the software complexity \cite{Akalanka2022}. The proposed system predicts the software complexity level through a supervised machine learning model. It also provides a tool to reduce the complexity of the source code by using a visual representation by applying simple refactoring techniques without user intervention: clear unused variables, unused methods, empty methods, unreachable code blocks, comments, and empty lines. Although these refactoring techniques make source code clearer, they do not decrease the SSCC of a method.

Existing approaches can indirectly reduce software cognitive complexity \cite{Haas2016,Hubert2019,Akalanka2022}. Nevertheless, they are not able to automatically reduce methods cognitive complexity to a given threshold.  In a previous work, we modelled the reduction of the SSCC to a given threshold as an optimization problem where the search space contains all feasible sequences of code extraction opportunities \cite{Saborido2022}. We also defined two algorithms which enumerate sequences of code extraction refactoring operations until a stopping criterion is met. However, enumeration algorithms fail to scale with the code size and they cannot assure that no solution exist. Our proposal here is novel because we model the software cognitive complexity reduction to a given threshold as an ILP optimization problem. This makes it feasible to apply efficient solvers, like CPLEX, to get optimal solutions very quickly.

\section{Conclusion}
\label{sec:conclusion}
We formulated the reduction of software cognitive complexity provided by SonarCloud and SonarQube, to a given threshold, as an ILP optimization problem. We then proposed an approach to automatically reduce the cognitive complexity of methods in software projects to the chosen threshold using CPLEX as solver. 
We conducted some experiments in 10 open-source software projects, analyzing 1,050 methods with a cognitive complexity greater than the default threshold suggested by SonarQube (15). The proposed approach was able to reduce the cognitive complexity to or below the threshold in 84\% of those methods, taking no more than five seconds on average. The cognitive complexity reduction, the number of extracted LOC, and the number of parameters of new methods, average 11.52, 24, and four, respectively.

As future work, we plan to validate our approach on software developers in order to get their feedback and analyze the way of including our approach as part of the continuous integration practice. Although it is not taken into account by the SSCC metric, the name of methods can influence the understanding of the source code. An aspect that is out of the scope of this article is the choice of the name for the new extracted methods. This is an important aspect we plan to address in the near future. Large language models based approaches, such as Chat-GPT or GitHub Copilot, could be used for this purpose.

\ifCLASSOPTIONcompsoc
  \section*{Acknowledgments}
\else
  \section*{Acknowledgment}
\fi
This research has been supported by Universidad de Málaga (grant B1-2020\_01). Rubén Saborido is a postdoctoral researcher funded by the Andalusian PAIDI program (POSTDOC\_21\_00567).

\bibliographystyle{IEEEtran}
\bibliography{bibliography}

\begin{thebibliography}{10}
\providecommand{\url}[1]{#1}
\csname url@samestyle\endcsname
\providecommand{\newblock}{\relax}
\providecommand{\bibinfo}[2]{#2}
\providecommand{\BIBentrySTDinterwordspacing}{\spaceskip=0pt\relax}
\providecommand{\BIBentryALTinterwordstretchfactor}{4}
\providecommand{\BIBentryALTinterwordspacing}{\spaceskip=\fontdimen2\font plus
\BIBentryALTinterwordstretchfactor\fontdimen3\font minus
  \fontdimen4\font\relax}
\providecommand{\BIBforeignlanguage}[2]{{%
\expandafter\ifx\csname l@#1\endcsname\relax
\typeout{** WARNING: IEEEtran.bst: No hyphenation pattern has been}%
\typeout{** loaded for the language `#1'. Using the pattern for}%
\typeout{** the default language instead.}%
\else
\language=\csname l@#1\endcsname
\fi
#2}}
\providecommand{\BIBdecl}{\relax}
\BIBdecl

\bibitem{Campbell2018}
\BIBentryALTinterwordspacing
G.~A. Campbell, ``Cognitive {Complexity}: {An} {Overview} and {Evaluation},''
  in \emph{Proceedings of the 2018 {International} {Conference} on {Technical}
  {Debt}}, ser. {TechDebt} '18.\hskip 1em plus 0.5em minus 0.4em\relax New
  York, NY, USA: Association for Computing Machinery, 2018, pp. 57--58.
  [Online]. Available: \url{https://doi.org/10.1145/3194164.3194186}
\BIBentrySTDinterwordspacing

\bibitem{Saborido2022}
R.~Saborido, J.~Ferrer, F.~Chicano, and E.~Alba, ``Automatizing software
  cognitive complexity reduction,'' \emph{IEEE Access}, vol.~10, pp.
  11\,642--11\,656, 2022.

\bibitem{Yin2002}
R.~K. Yin, \emph{Case Study Research: Design and Methods - Third Edition},
  3rd~ed.\hskip 1em plus 0.5em minus 0.4em\relax SAGE Publications, 2002.

\bibitem{McCabe1976}
T.~J. McCabe, ``A complexity measure,'' \emph{IEEE Trans. Softw. Eng.}, vol.~2,
  no.~4, p. 308–320, Jul. 1976.

\bibitem{Halstead1977}
M.~H. Halstead, \emph{\BIBforeignlanguage{English}{Elements of software science
  / Maurice H. Halstead}}.\hskip 1em plus 0.5em minus 0.4em\relax Elsevier New
  York, 1977.

\bibitem{Woodward1979}
M.~Woodward, M.~Hennell, and D.~Hedley, ``A measure of control flow complexity
  in program text,'' \emph{IEEE Transactions on Software Engineering}, vol.
  SE-5, no.~1, pp. 45--50, Jan 1979.

\bibitem{DBLP:conf/ppig/DouceLB99}
C.~R. Douce, P.~J. Layzell, and J.~Buckley, ``Spatial measures of software
  complexity,'' in \emph{{PPIG}}.\hskip 1em plus 0.5em minus 0.4em\relax
  Psychology of Programming Interest Group, 1999, p.~6.

\bibitem{Misra2007b}
S.~{Misra}, ``An object oriented complexity metric based on cognitive
  weights,'' in \emph{6th IEEE International Conference on Cognitive
  Informatics}, 2007, pp. 134--139.

\bibitem{Misra2008}
S.~Misra and I.~Akman, ``A model for measuring cognitive complexity of
  software,'' in \emph{Knowledge-Based Intelligent Information and Engineering
  Systems}, I.~Lovrek, R.~J. Howlett, and L.~C. Jain, Eds.\hskip 1em plus 0.5em
  minus 0.4em\relax Berlin, Heidelberg: Springer Berlin Heidelberg, 2008, pp.
  879--886.

\bibitem{Misra2008b}
S.~{Misra} and I.~Akman, ``A new complexity metric based on cognitive
  informatics,'' in \emph{Rough Sets and Knowledge Technology}, G.~Wang, T.~Li,
  J.~W. Grzymala-Busse, D.~Miao, A.~Skowron, and Y.~Yao, Eds.\hskip 1em plus
  0.5em minus 0.4em\relax Berlin, Heidelberg: Springer Berlin Heidelberg, 2008,
  pp. 620--627.

\bibitem{MISRA2011}
\BIBentryALTinterwordspacing
S.~Misra, I.~Akman, and M.~Koyuncu, ``An inheritance complexity metric for
  object-oriented code: A cognitive approach,'' \emph{Sadhana}, vol.~36, no.~3,
  p. 317, Jul 2011. [Online]. Available:
  \url{https://doi.org/10.1007/s12046-011-0028-2}
\BIBentrySTDinterwordspacing

\bibitem{misra_suite_2012}
S.~Misra, M.~Koyuncu, M.~Crasso, C.~Mateos, and A.~Zunino, ``A {Suite} of
  {Cognitive} {Complexity} {Metrics},'' in \emph{Computational {Science} and
  {Its} {Applications} – {ICCSA} 2012}, B.~Murgante, O.~Gervasi, S.~Misra,
  N.~Nedjah, A.~M. A.~C. Rocha, D.~Taniar, and B.~O. Apduhan, Eds.\hskip 1em
  plus 0.5em minus 0.4em\relax Berlin, Heidelberg: Springer Berlin Heidelberg,
  2012, pp. 234--247.

\bibitem{iet-sen.2011.0206}
\BIBentryALTinterwordspacing
S.~Misra, ``\BIBforeignlanguage{English}{Framework for evaluation and
  validation of software complexity measures},''
  \emph{\BIBforeignlanguage{English}{IET Software}}, vol.~6, pp. 323--334(11),
  August 2012. [Online]. Available:
  \url{https://digital-library.theiet.org/content/journals/10.1049/iet-sen.2011.0206}
\BIBentrySTDinterwordspacing

\bibitem{Haas2016}
R.~Haas and B.~Hummel, ``Deriving {Extract} {Method} {Refactoring}
  {Suggestions} for {Long} {Methods},'' in \emph{Software {Quality}. {The}
  {Future} of {Systems}- and {Software} {Development}}, D.~Winkler, S.~Biffl,
  and J.~Bergsmann, Eds.\hskip 1em plus 0.5em minus 0.4em\relax Cham: Springer
  International Publishing, 2016, pp. 144--155.

\bibitem{wijendra_automated_2016}
D.~R. Wijendra and K.~P. Hewagamage, ``Automated tool for the calculation of
  cognitive complexity of a software,'' in \emph{2016 2nd {International}
  {Conference} on {Science} in {Information} {Technology} ({ICSITech})}, 2016,
  pp. 163--168.

\bibitem{crasso_assessing_2016}
\BIBentryALTinterwordspacing
M.~Crasso, C.~Mateos, A.~Zunino, S.~Misra, and P.~Polvorín, ``Assessing
  {Cognitive} {Complexity} in {Java}-{Based} {Object}-{Oriented} {Systems}:
  {Metrics} and {Tool} {Support},'' \emph{COMPUTING AND INFORMATICS}, vol.~35,
  no.~3, pp. 497--527, Nov. 2016. [Online]. Available:
  \url{http://www.cai.sk/ojs/index.php/cai/article/view/1747}
\BIBentrySTDinterwordspacing

\bibitem{kaur_cognitive_2019}
\BIBentryALTinterwordspacing
L.~Kaur and A.~Mishra, ``Cognitive complexity as a quantifier of version to
  version {Java}-based source code change: {An} empirical probe,''
  \emph{Information and Software Technology}, vol. 106, pp. 31--48, 2019.
  [Online]. Available:
  \url{https://www.sciencedirect.com/science/article/pii/S0950584918301903}
\BIBentrySTDinterwordspacing

\bibitem{alqadi_relationship_2019}
B.~S. Alqadi, ``The {Relationship} {Between} {Cognitive} {Complexity} and the
  {Probability} of {Defects},'' in \emph{2019 {IEEE} {International}
  {Conference} on {Software} {Maintenance} and {Evolution} ({ICSME})}, 2019,
  pp. 600--604.

\bibitem{Hubert2019}
J.~Hubert, ``\BIBforeignlanguage{English}{Implementation of an automatic
  extract method refactoring},'' Master's thesis, University of Stuttgart,
  Faculty of Computer Science, Electrical Engineering, and Information
  Technology, Germany, Apr. 2019.

\bibitem{munoz_baron_empirical_2020}
\BIBentryALTinterwordspacing
M.~Muñoz~Barón, M.~Wyrich, and S.~Wagner, ``An {Empirical} {Validation} of
  {Cognitive} {Complexity} as a {Measure} of {Source} {Code}
  {Understandability},'' in \emph{Proceedings of the 14th {ACM} / {IEEE}
  {International} {Symposium} on {Empirical} {Software} {Engineering} and
  {Measurement} ({ESEM})}, ser. {ESEM} '20.\hskip 1em plus 0.5em minus
  0.4em\relax New York, NY, USA: Association for Computing Machinery, 2020,
  event-place: Bari, Italy. [Online]. Available:
  \url{https://doi.org/10.1145/3382494.3410636}
\BIBentrySTDinterwordspacing

\bibitem{Wijendra2022}
D.~R. Wijendra and K.~P. Hewagamage, ``Cognitive complexity reduction through
  control flow graph generation,'' in \emph{2022 IEEE 7th International
  conference for Convergence in Technology (I2CT)}, 2022, pp. 1--7.

\bibitem{Wijendra2022b}
------, ``Application of the refactoring to the understandability and the
  cognitive complexity of a software,'' in \emph{2022 IEEE 7th International
  conference for Convergence in Technology (I2CT)}, 2022, pp. 1--6.

\bibitem{Akalanka2022}
M.~Akalanka, W.~Weerasinghe, H.~Perera, T.~Kumari, D.~Wijendra, and
  J.~Krishara, ``Software complexity automation tool for industrial practices
  with qualitative and quantitative aspects,'' in \emph{2022 4th International
  Conference on Advancements in Computing (ICAC)}, 2022, pp. 453--458.

\end{thebibliography}

%-\input{biography}

\newpage
\clearpage

\pagenumbering{roman}
\onecolumn
\appendix[Automatizing Software Cognitive Complexity Reduction through ILP]

''In order to maintain the clarity and conciseness of the main body of this research paper, certain figures that provide additional context and support to the research findings have been excluded due to space limitations. These figures are presented in this appendix to ensure that readers have access to the complete set of information related to the study. Each figure is labeled with a number and accompanied by a descriptive title. Please refer to the corresponding figure numbers mentioned in the main text for a comprehensive understanding of the research findings. It is important to note that these figures are supplemental and not essential to the main argument of the paper. The decision to exclude them was made to prioritize the flow and readability of the main body."
\\

\noindent
\emph{The previous paragraph was generated by an AI tool.}

\renewcommand{\thefigure}{A.\arabic{figure}}
\setcounter{figure}{0}

\renewcommand{\thetable}{A.\arabic{table}}
\setcounter{table}{0}

\renewcommand{\theequation}{\hbAppendixPrefix\arabic{equation}} 
\setcounter{equation}{0}

\begin{figure*}[h!]
\begin{lstlisting}[numbers=left,xleftmargin=+0.6cm,xrightmargin=+0.2cm,frame=single,framexleftmargin=0em,escapechar=ñ,stepnumber=1,
    showstringspaces=false,
    tabsize=1,
    breaklines=true,
    breakatwhitespace=false,
    basicstyle=\tiny,
    language=,
    mathescape=true]{}
\ENCODING=ISO-8859-1
\Problem name: ilog.cplex

Minimize
 obj1: X4 + X5 + X7 + X6 + X2 + X3 + X8 + X9 + X10 + X1 + X0
Subject To
 c1:  X4 + X5 <= 1
 c2:  X4 + X7 <= 1
 c3:  X7 + X6 <= 1
 c4:  Z20 - X2 <= 0
 c5:  X2 + 2 Z30 - X3 <= 1
 c6:  - X4 + X2 + X3 + 3 Z40 <= 2
 c7:  X4 + X5 - X6 + X2 + X3 + 5 Z60 <= 4
 c8:  X4 + X5 + X7 + X6 + X2 + X3 + 7 Z80 - X8 <= 6
 c9:  X4 + X5 + X7 + X6 + X2 + X3 + X8 + 8 Z90 - X9 <= 7
 c10: - X5 + X2 + X3 + 3 Z50 <= 2
 c11: X5 - X7 + X2 + X3 + 4 Z70 <= 3
 c12: X5 + X7 + X2 + X3 + 5 Z100 - X10 <= 4
 c13: Z10 - X1 <= 0
 c14: - 15 Z20 - 12 Z30 - 10 Z40 - 10 Z60 - 10 Z80 - 7 Z90 - 12 Z50 - 12 Z70 - 2 Z100 - Z10 + X0 <= 0
 c15: - 14 X1 <= 0
 c16: - X3 + Z32 <= 0
 c17: - X4 + X3 + 2 Z42 <= 1
 c18: X4 + X5 - X6 + X3 + 4 Z62 <= 3
 c19: X4 + X5 + X7 + X6 + X3 - X8 + 6 Z82 <= 5
 c20: X4 + X5 + X7 + X6 + X3 + X8 - X9 + 7 Z92 <= 6
 c21: - X5 + X3 + 2 Z52 <= 1
 c22: X5 - X7 + X3 + 3 Z72 <= 2
 c23: X5 + X7 + X3 - X10 + 4 Z102 <= 3
 c24: - 12 Z32 - 10 Z42 - 10 Z62 - 10 Z82 - 7 Z92 - 12 Z52 - 12 Z72 - 2 Z102 <= 0
 c25: - X4 + Z43 <= 0
 c26: X4 + X5 - X6 + 3 Z63 <= 2
 c27: X4 + X5 + X7 + X6 - X8 + 5 Z83 <= 4
 c28: X4 + X5 + X7 + X6 + X8 - X9 + 6 Z93 <= 5
 c29: - X5 + Z53 <= 0
 c30: X5 - X7 + 2 Z73 <= 1
 c31: X5 + X7 - X10 + 3 Z103 <= 2
 c32: - 7 X3 - 7 Z43 - 7 Z63 - 7 Z83 - 5 Z93 - 8 Z53 - 8 Z73 - Z103 <= 0
 c33: - X6 + Z64 <= 0
 c34: X5 + X7 + X6 + X2 + X3 - X8 + X0 + 7 Z84 <= 6
 c35: X5 + X7 + X6 + X2 + X3 + X8 - X9 + X0 + 8 Z94 <= 7
 c36: - 8 X4 - 7 Z64 - 7 Z84 - 5 Z94 <= 0
 c37: X4 - X6 + X2 + X3 + X0 + 5 Z65 <= 4
 c38: X4 + X7 + X6 + X2 + X3 - X8 + X0 + 7 Z85 <= 6
 c39: X4 + X7 + X6 + X2 + X3 + X8 - X9 + X0 + 8 Z95 <= 7
 c40: - X7 + Z75 <= 0
 c41: X7 - X10 + 2 Z105 <= 1
 c42: - 7 X5 - 7 Z65 - 7 Z85 - 5 Z95 - 8 Z75 - Z105 <= 0
 c43: - X8 + Z86 <= 0
 c44: X8 - X9 + 2 Z96 <= 1
 c45: - 8 X6 - 7 Z86 - 5 Z96 <= 0
 c46: X4 + X5 + X6 + X2 + X3 - X8 + X0 + 7 Z87 <= 6
 c47: X4 + X5 + X6 + X2 + X3 + X8 - X9 + X0 + 8 Z97 <= 7
 c48: - X10 + Z107 <= 0
 c49: - 7 X7 - 7 Z87 - 5 Z97 - Z107 <= 0
 c50: - X9 + Z98 <= 0
 c51: - 8 X8 - 5 Z98 <= 0
 c52: - 12 X9 <= 0
 c53: - 14 X10 <= 0
 c54: X0  = 1
Bounds
 0 <= X4 <= 1
 0 <= X5 <= 1
 0 <= X7 <= 1
 0 <= X6 <= 1
 0 <= Z20 <= 1
 0 <= X2 <= 1
 0 <= Z30 <= 1
 0 <= X3 <= 1
 0 <= Z40 <= 1
 0 <= Z60 <= 1
 0 <= Z80 <= 1
 0 <= X8 <= 1
 0 <= Z90 <= 1
 0 <= X9 <= 1
 0 <= Z50 <= 1
 0 <= Z70 <= 1
 0 <= Z100 <= 1
 0 <= X10 <= 1
 0 <= Z10 <= 1
 0 <= X1 <= 1
 0 <= X0 <= 1
 0 <= Z32 <= 1
 0 <= Z42 <= 1
 0 <= Z62 <= 1
 0 <= Z82 <= 1
 0 <= Z92 <= 1
 0 <= Z52 <= 1
 0 <= Z72 <= 1
 0 <= Z102 <= 1
 0 <= Z43 <= 1
 0 <= Z63 <= 1
 0 <= Z83 <= 1
 0 <= Z93 <= 1
 0 <= Z53 <= 1
 0 <= Z73 <= 1
 0 <= Z103 <= 1
 0 <= Z64 <= 1
 0 <= Z84 <= 1
 0 <= Z94 <= 1
 0 <= Z65 <= 1
 0 <= Z85 <= 1
 0 <= Z95 <= 1
 0 <= Z75 <= 1
 0 <= Z105 <= 1
 0 <= Z86 <= 1
 0 <= Z96 <= 1
 0 <= Z87 <= 1
 0 <= Z97 <= 1
 0 <= Z107 <= 1
 0 <= Z98 <= 1
Binaries
 X4  X5  X7  X6  Z20  X2  Z30  X3  Z40  Z60  Z80 X8  Z90  X9  Z50  Z70  Z100  X10  Z10  X1  X0  Z32  Z42  Z62  Z82  Z92  Z52  Z72  Z102  Z43  Z63  Z83  Z93 Z53  Z73  Z103  Z64  Z84  Z94  Z65  Z85  Z95  Z75  Z105  Z86  Z96  Z87  Z97 Z107  Z98 
End
\end{lstlisting}
\caption{CPLEX model of the ILP problem associated to the \texttt{hook} method in the \texttt{EZInjection} class of the open-source project ByteCode.}
\label{listingILPModelIllustratingExample}
\end{figure*}

%HISTOGRAMS

\begin{figure*}
    \centering
    \includegraphics[scale=0.5]{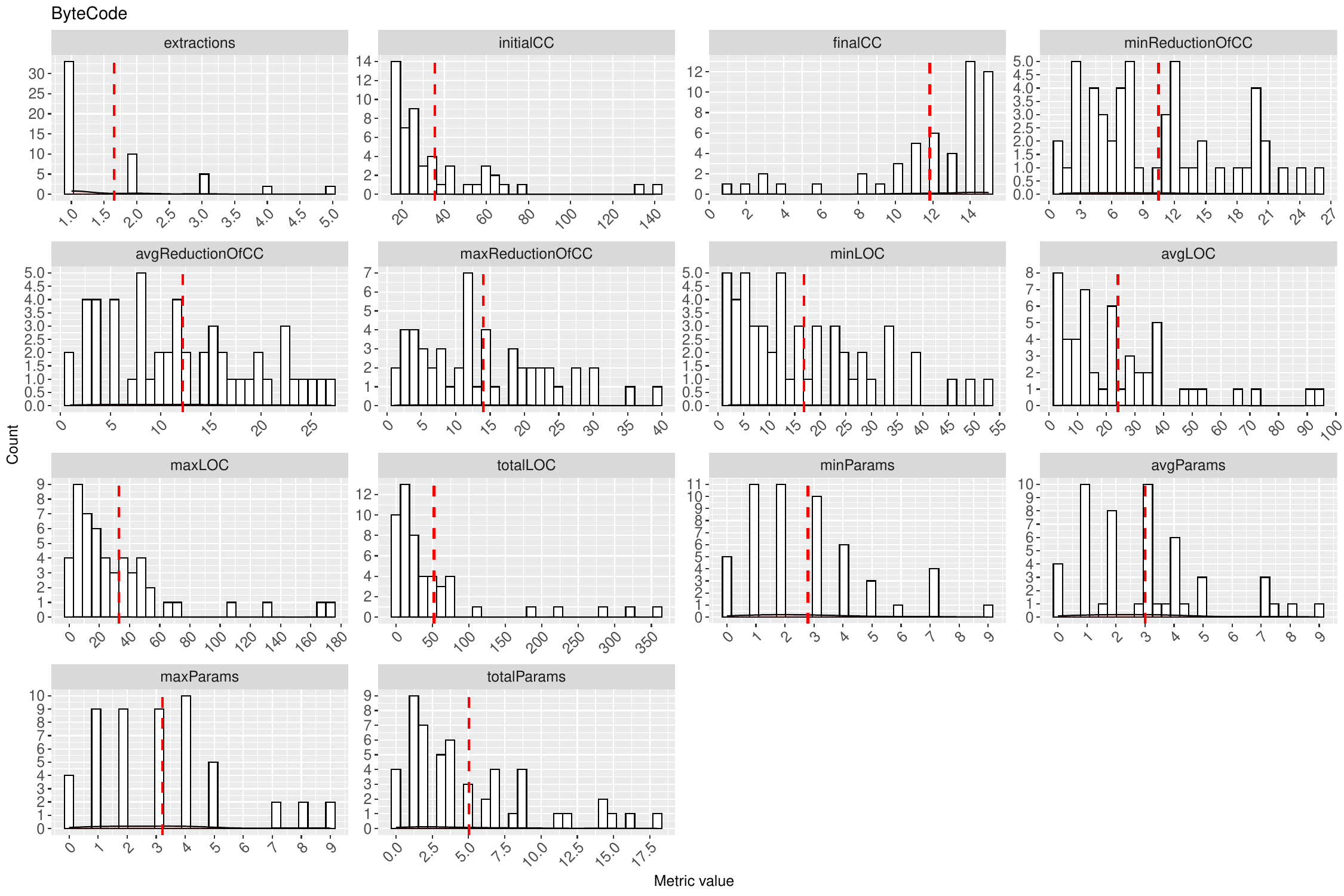}
    \caption{Frequency distribution of studied metrics for the 888 methods for which CPLEX found optimal solutions when using 300 seconds as stopping criterion for the open-source project ByteCode.}
    \label{fig:Histogram-ByteCode}
\end{figure*}

\begin{figure*}
    \centering
    \includegraphics[scale=0.5]{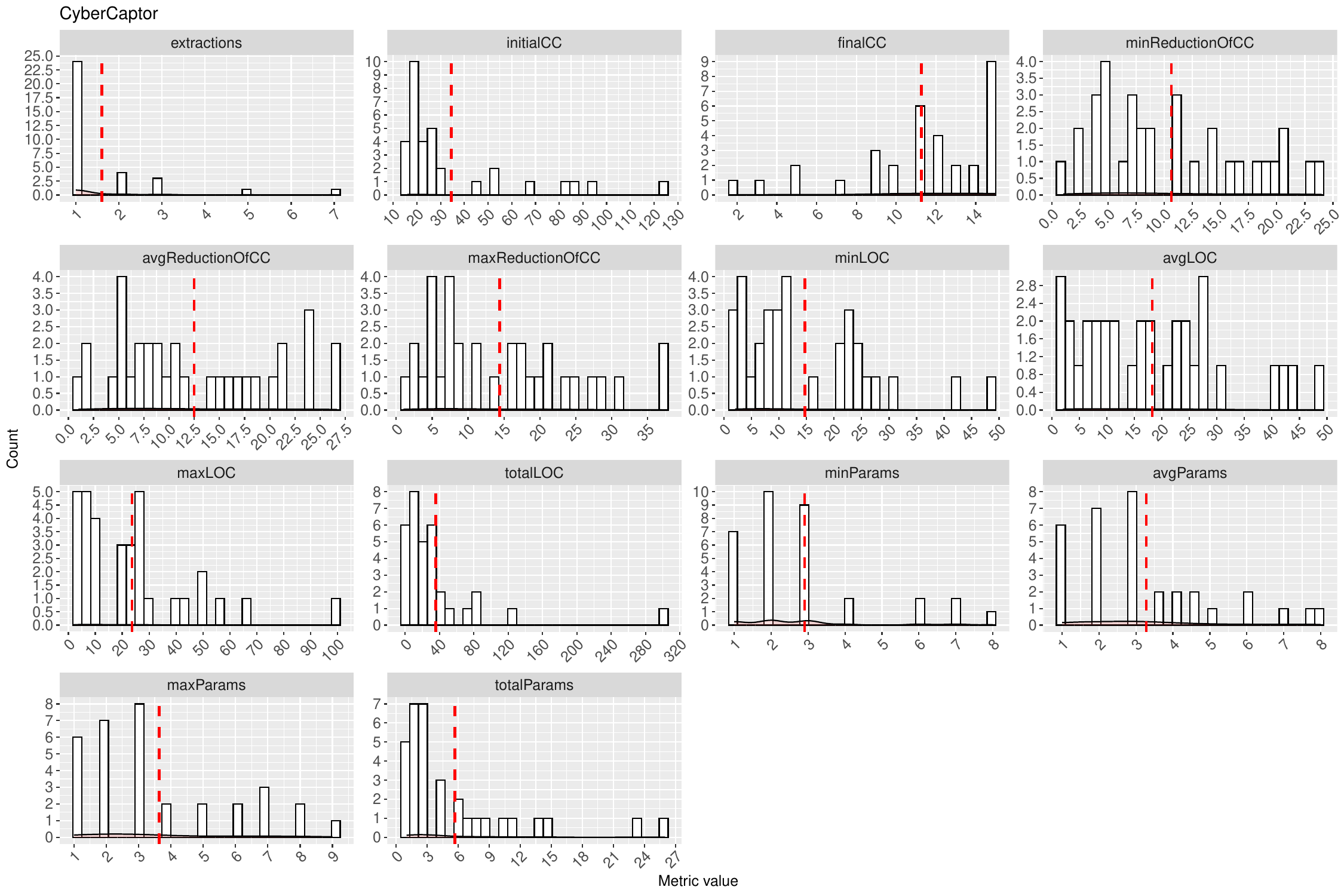}
    \caption{Frequency distribution of studied metrics for the 888 methods for which CPLEX found optimal solutions when using 300 seconds as stopping criterion for the open-source project CyberCaptor.}
    \label{fig:Histogram-CyberCaptor}
\end{figure*}

\begin{figure*}
    \centering
    \includegraphics[scale=0.5]{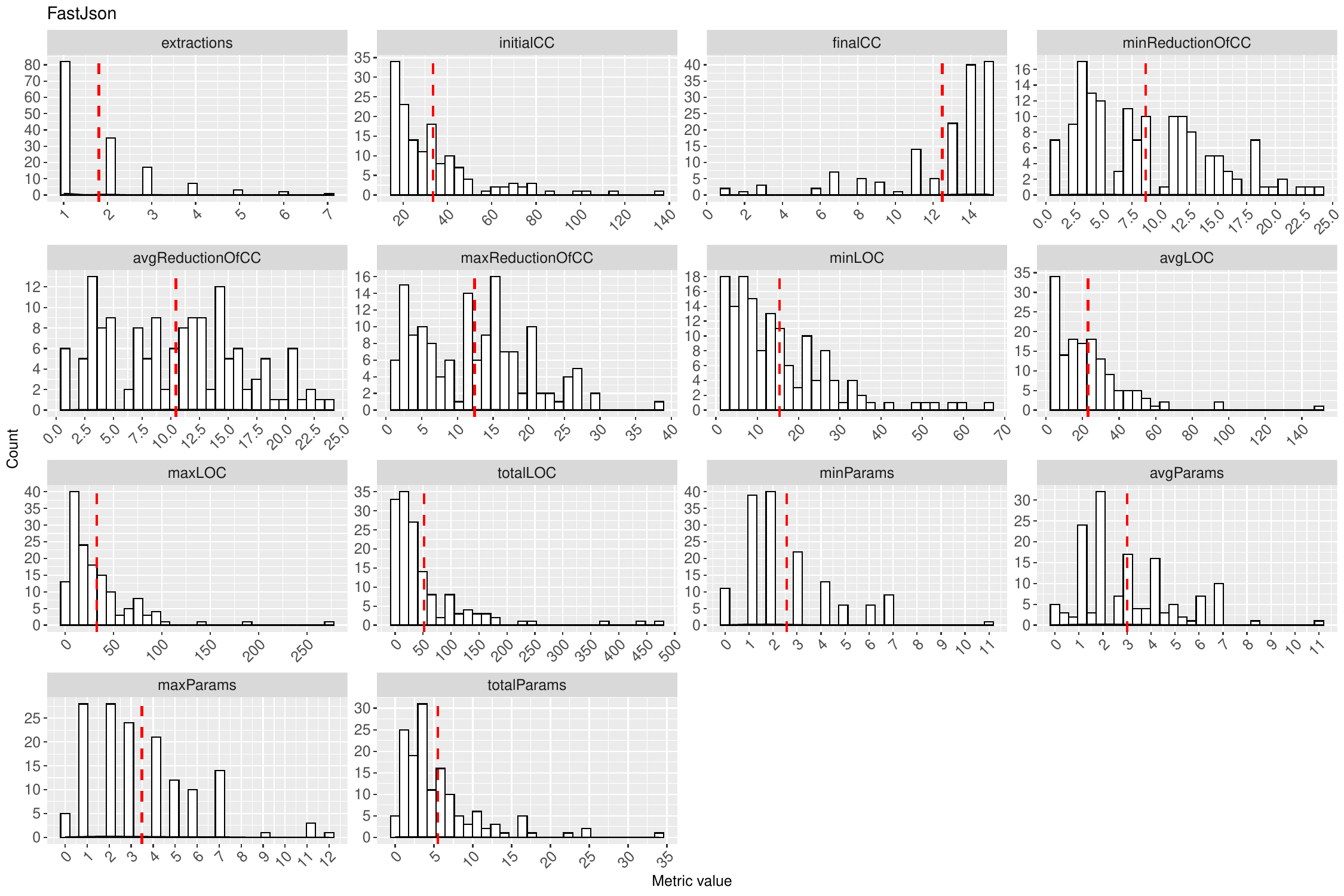}
    \caption{Frequency distribution of studied metrics for the 888 methods for which CPLEX found optimal solutions when using 300 seconds as stopping criterion for the open-source project FastJson.}
    \label{fig:Histogram-FastJson}
\end{figure*}

\begin{figure*}
    \centering
    \includegraphics[scale=0.5]{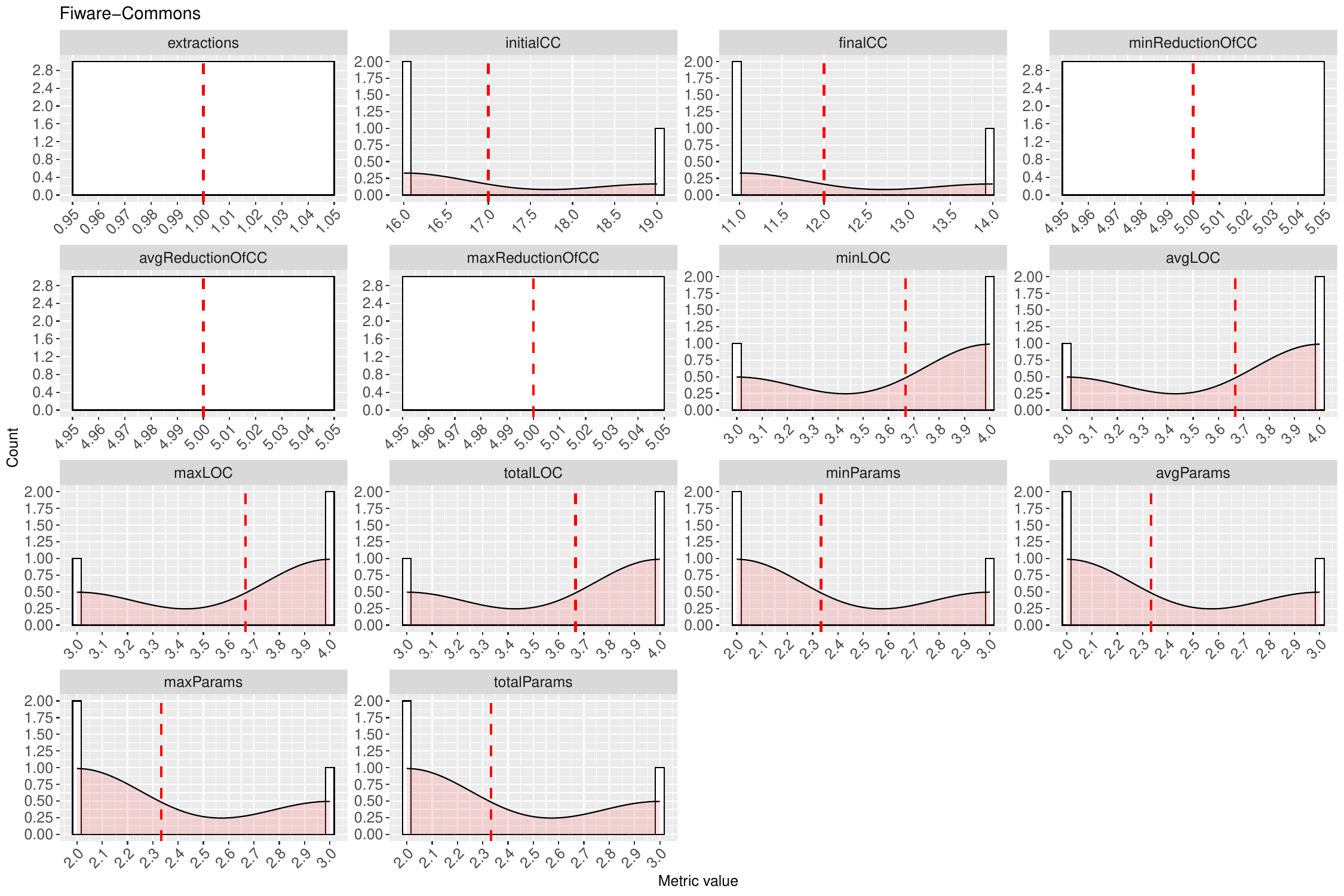}
    \caption{Frequency distribution of studied metrics for the 888 methods for which CPLEX found optimal solutions when using 300 seconds as stopping criterion for the open-source project Fiware-Commons.}
    \label{fig:Histogram-Fiware-Commons}
\end{figure*}

\begin{figure*}
    \centering
    \includegraphics[scale=0.5]{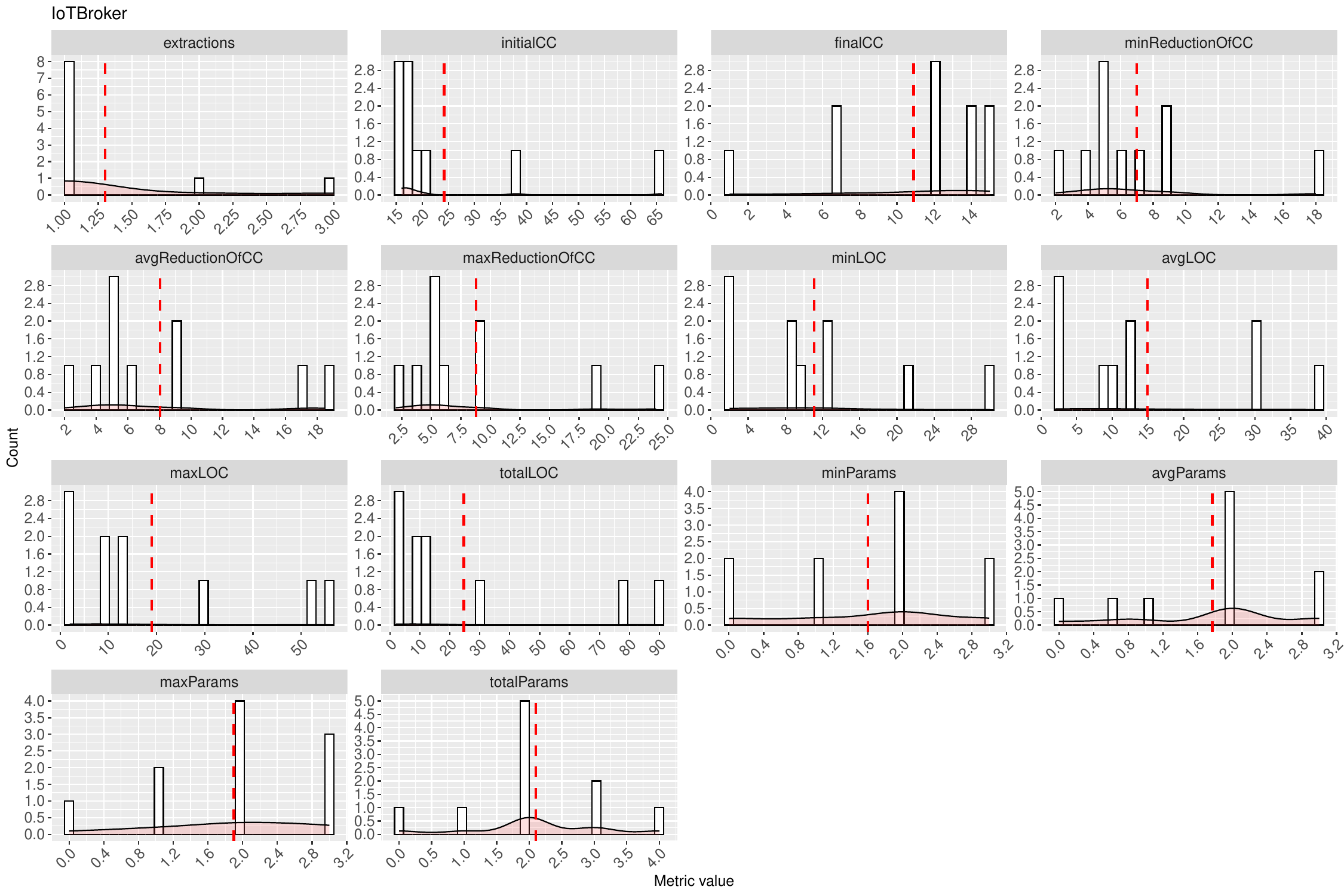}
    \caption{Frequency distribution of studied metrics for the 888 methods for which CPLEX found optimal solutions when using 300 seconds as stopping criterion for the open-source project IoTBroker.}
    \label{fig:Histogram-IoTBroker}
\end{figure*}

\begin{figure*}
    \centering
    \includegraphics[scale=0.5]{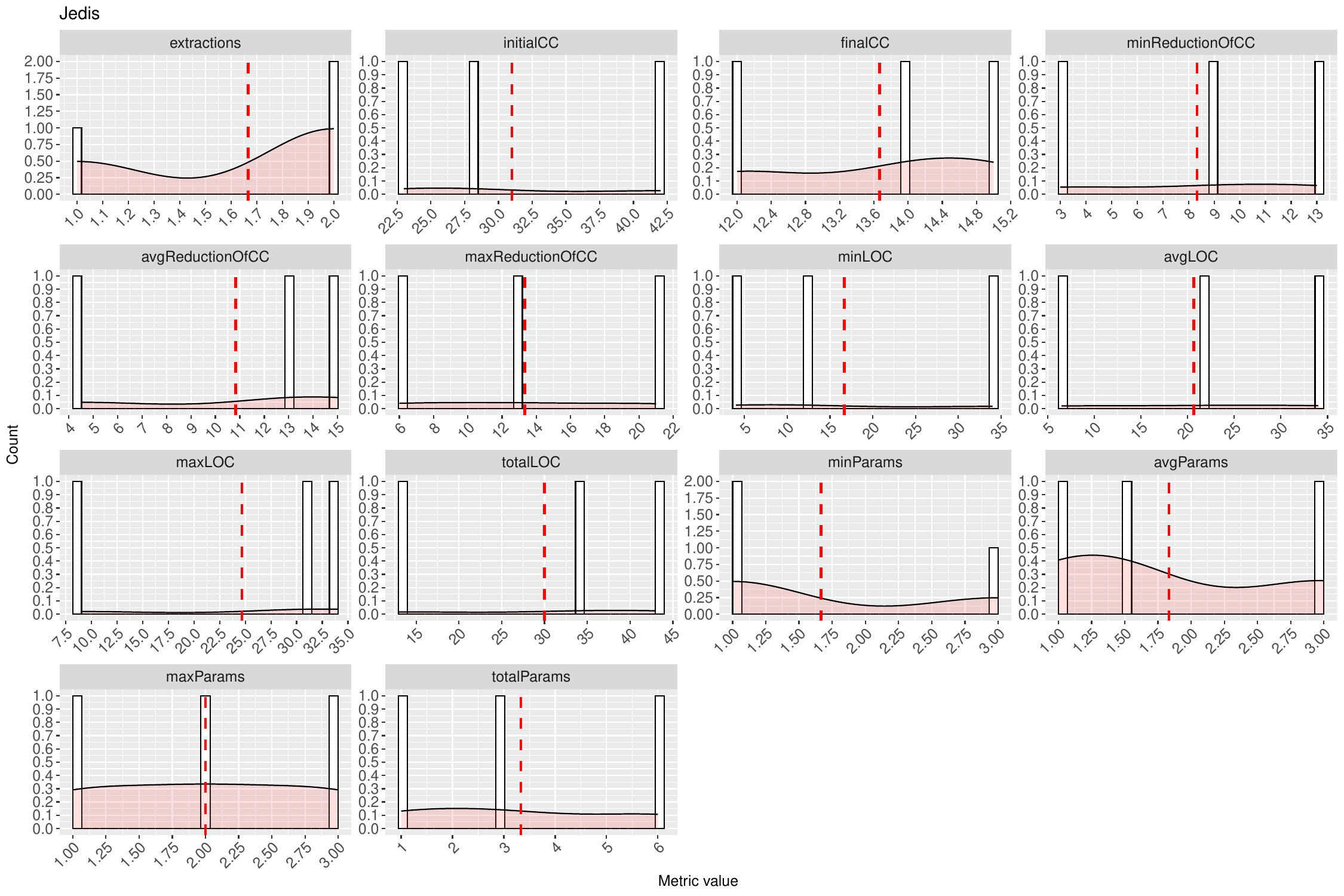}
    \caption{Frequency distribution of studied metrics for the 888 methods for which CPLEX found optimal solutions when using 300 seconds as stopping criterion for the open-source project Jedis.}
    \label{fig:Histogram-Jedis}
\end{figure*}

\begin{figure*}
    \centering
    \includegraphics[scale=0.5]{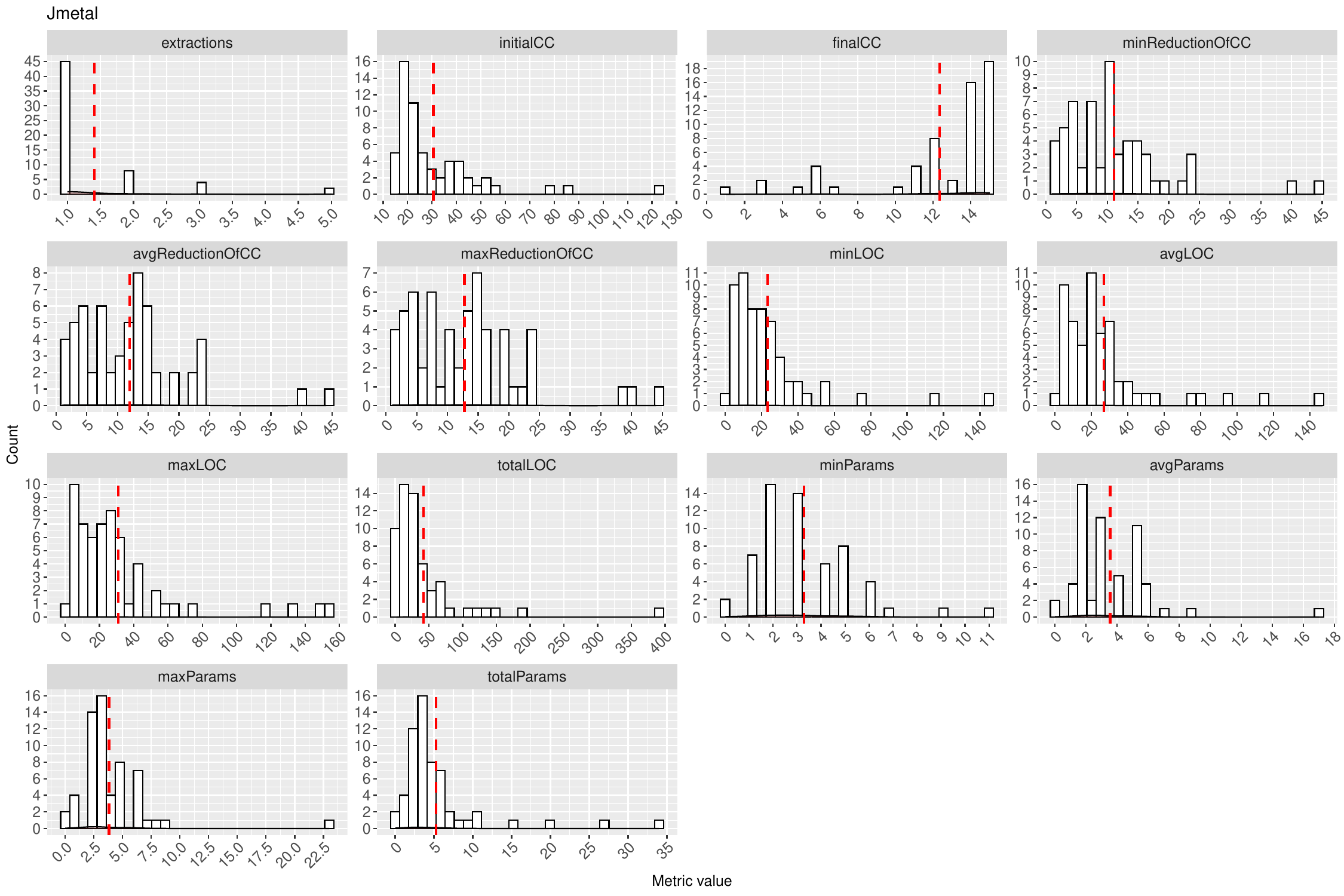}
    \caption{Frequency distribution of studied metrics for the 888 methods for which CPLEX found optimal solutions when using 300 seconds as stopping criterion for the open-source project jMetal.}
    \label{fig:Histogram-Jmetal}
\end{figure*}

\begin{figure*}
    \centering
    \includegraphics[scale=0.5]{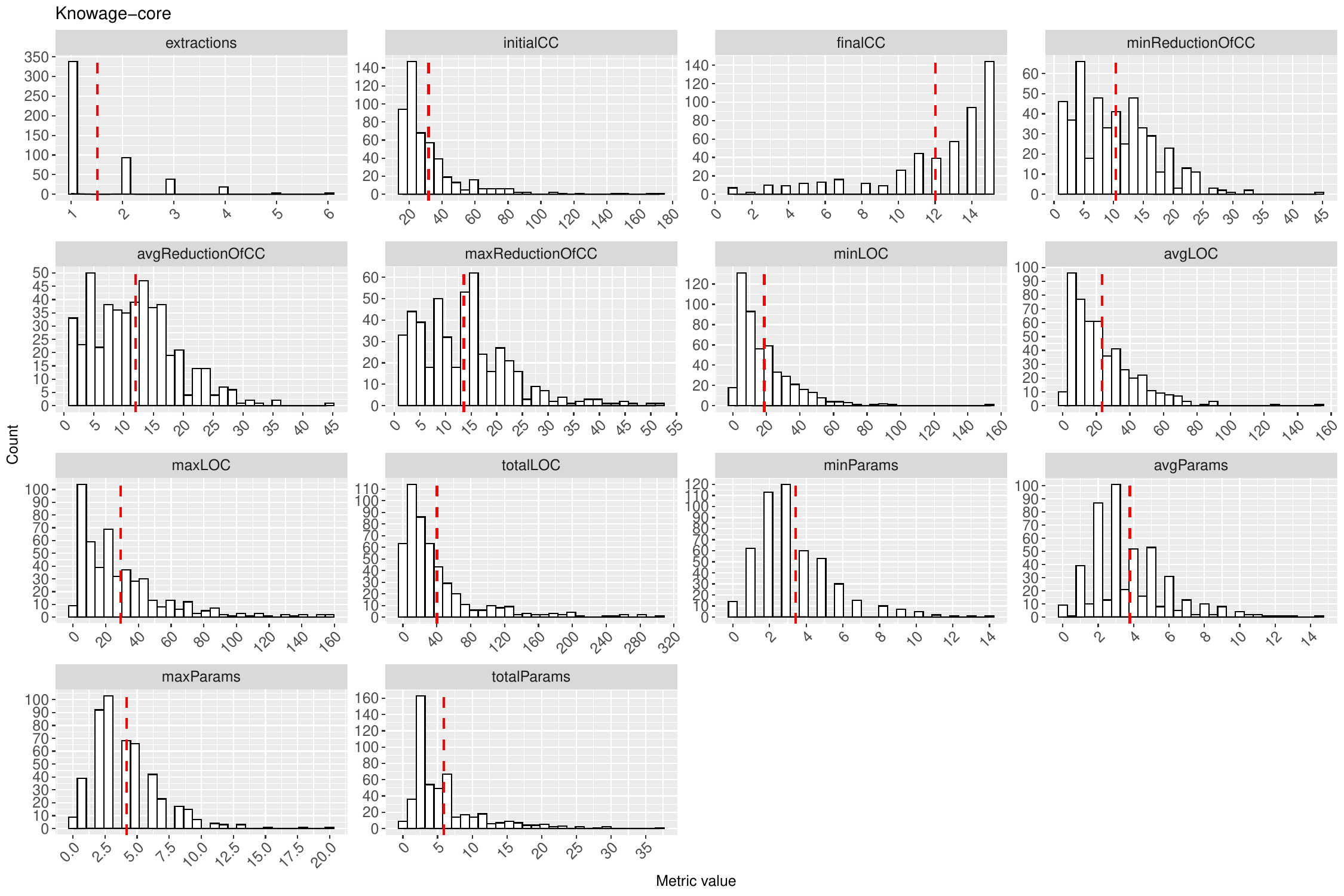}
    \caption{Frequency distribution of studied metrics for the 888 methods for which CPLEX found optimal solutions when using 300 seconds as stopping criterion for the open-source project Knowage-core.}
    \label{fig:Histogram-Knowage-core}
\end{figure*}

\begin{figure*}
    \centering
    \includegraphics[scale=0.5]{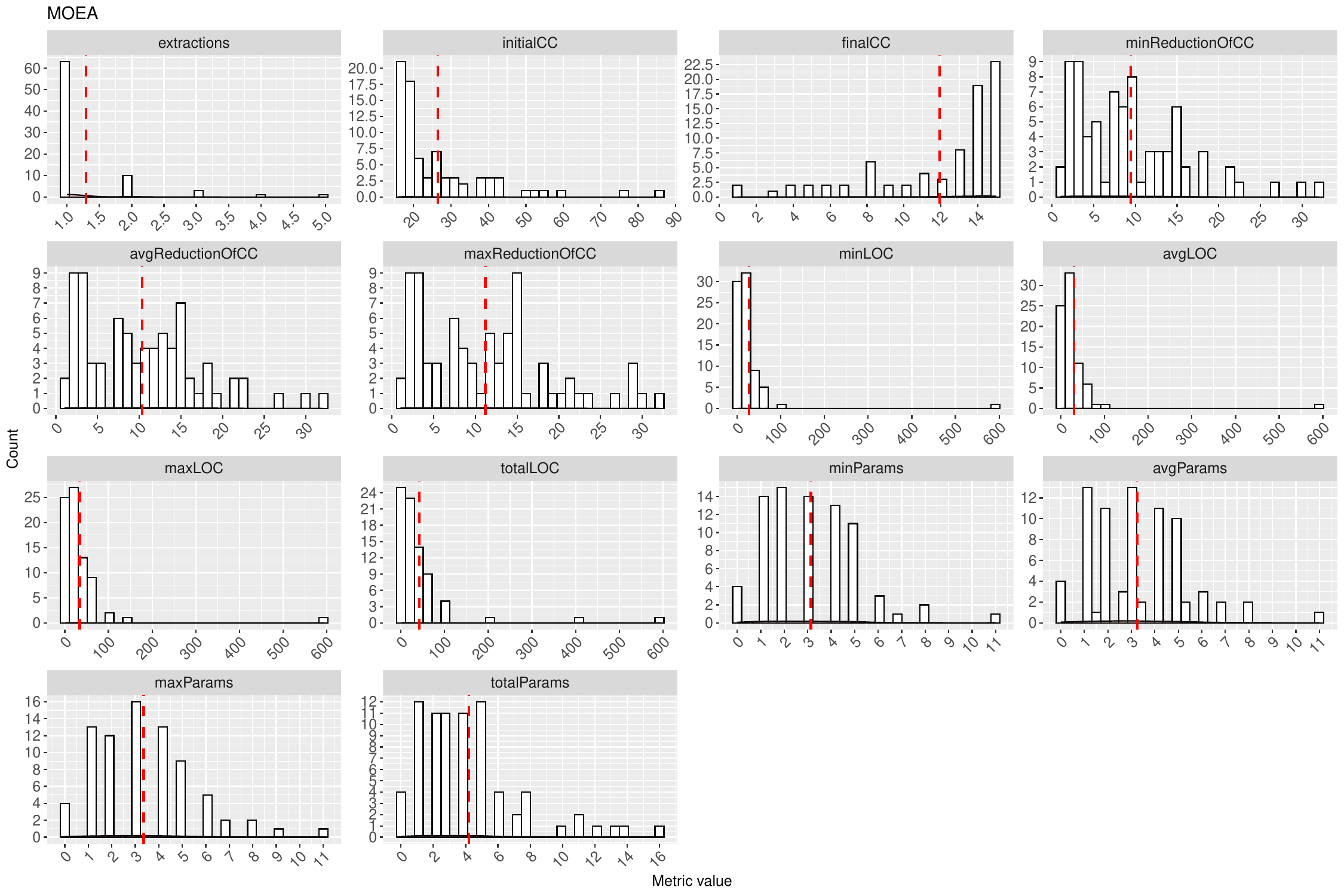}
    \caption{Frequency distribution of studied metrics for the 888 methods for which CPLEX found optimal solutions when using 300 seconds as stopping criterion for the open-source project MOEA-framework.}
    \label{fig:Histogram-MOEA}
\end{figure*}

\begin{figure*}
    \centering
    \includegraphics[scale=0.5]{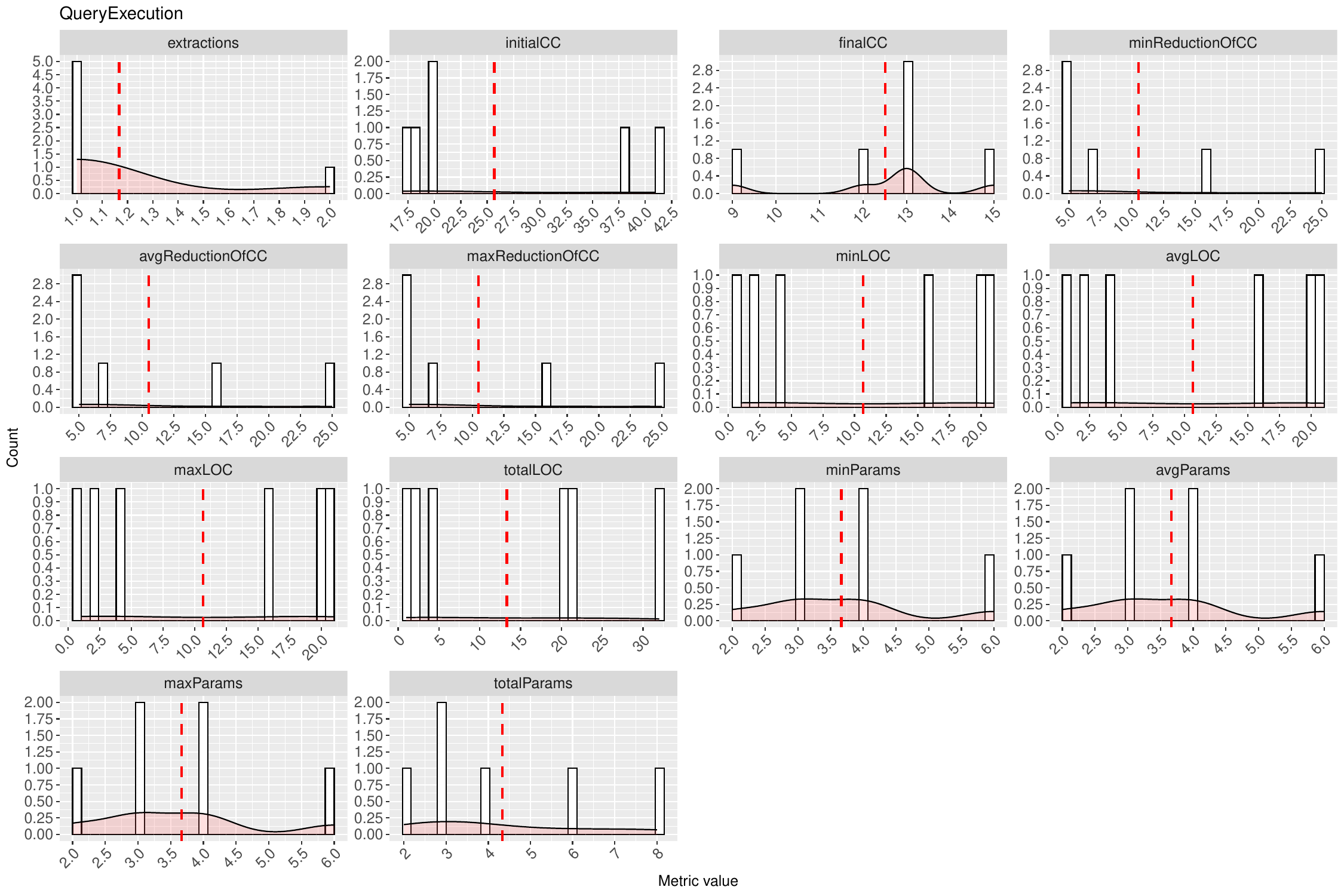}
    \caption{Frequency distribution of studied metrics for the 888 methods for which CPLEX found optimal solutions when using 300 seconds as stopping criterion for the open-source project QueryExecution.}
    \label{fig:Histogram-QueryExecution}
\end{figure*}

\begin{figure*}
\includegraphics[scale=0.48]{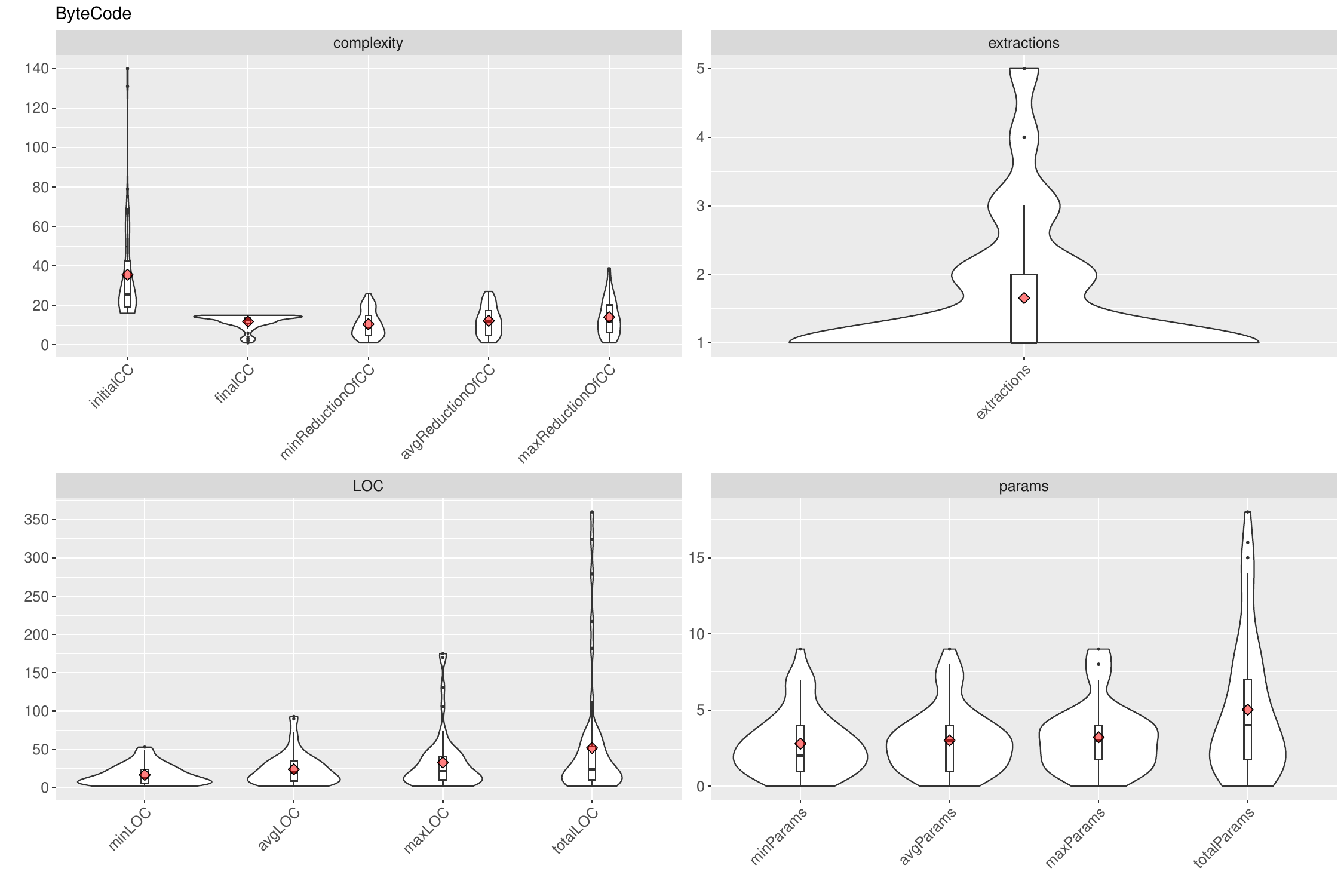}
    \caption{Summary statistics and the density of studied metrics for the 888 methods for which CPLEX found optimal solutions when using 300 seconds as stopping criterion though violin plots for the open-source project ByteCode.}
    \label{fig:ViolinplotByMetricsCategory-ByteCode}
\end{figure*}

\begin{figure*}
\includegraphics[scale=0.48]{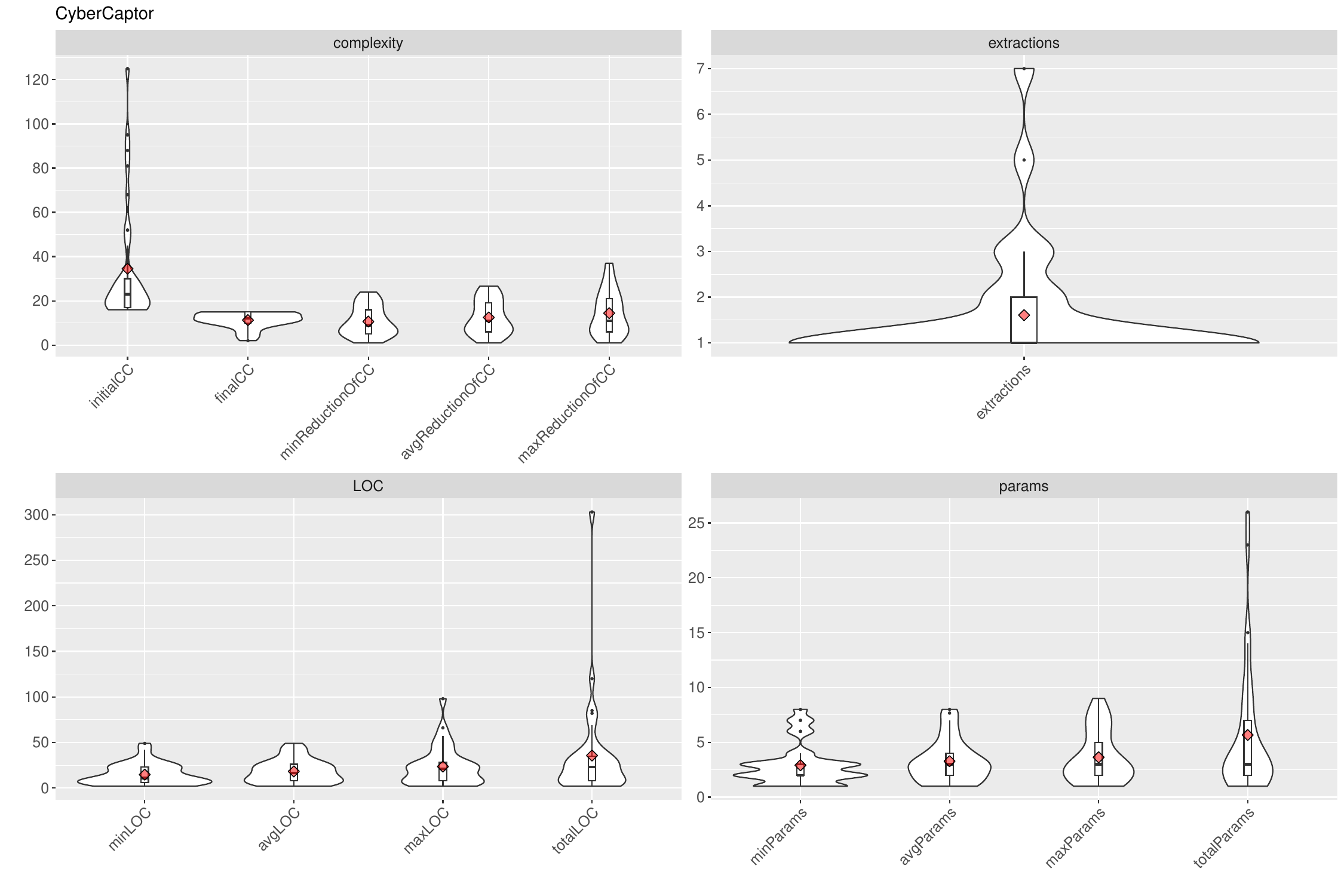}
    \caption{Summary statistics and the density of studied metrics for the 888 methods for which CPLEX found optimal solutions when using 300 seconds as stopping criterion though violin plots for the open-source project CyberCaptor.}
    \label{fig:ViolinplotByMetricsCategory-CyberCaptor}
\end{figure*}

\begin{figure*}
\includegraphics[scale=0.48]{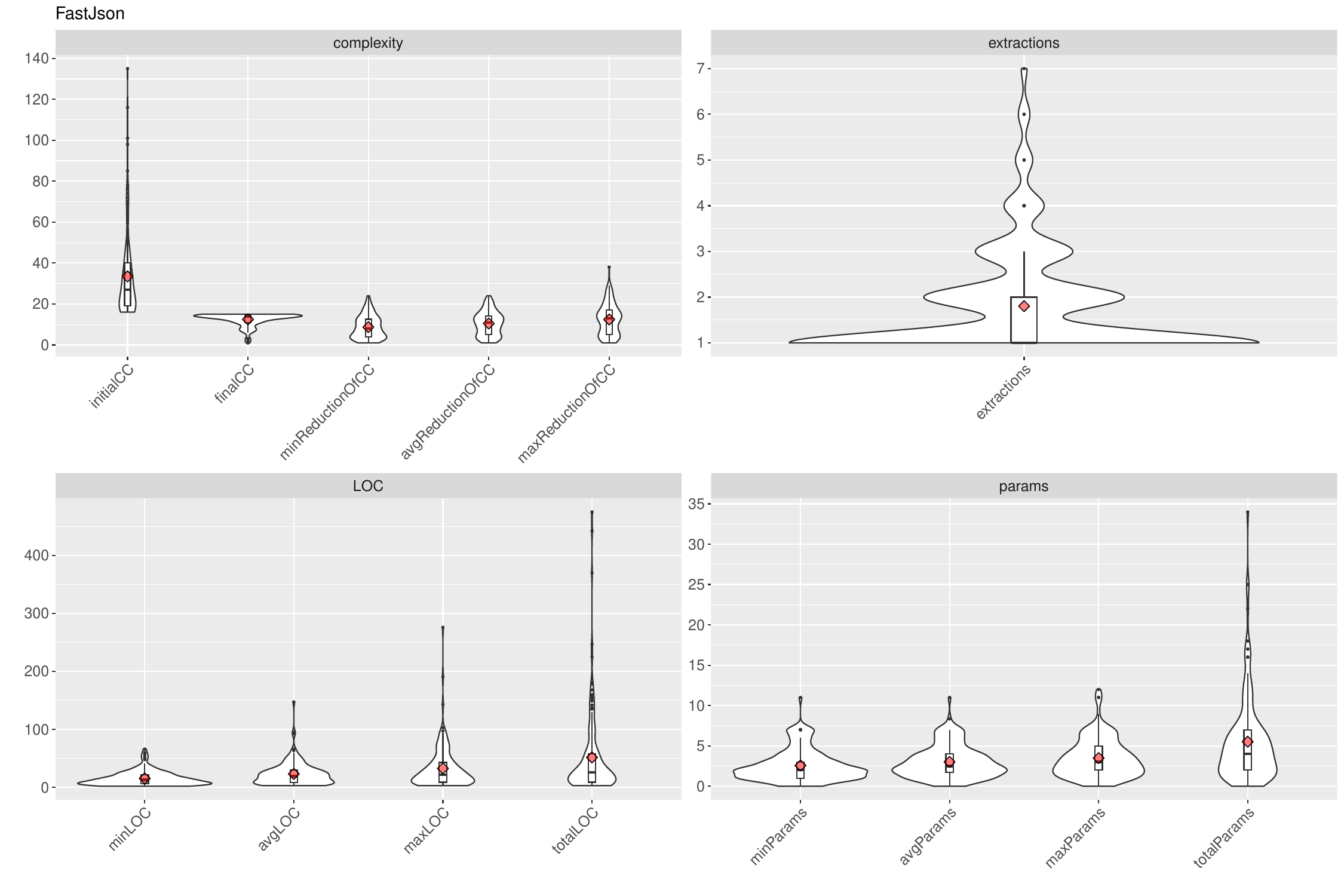}
    \caption{Summary statistics and the density of studied metrics for the 888 methods for which CPLEX found optimal solutions when using 300 seconds as stopping criterion though violin plots for the open-source project FastJson.}
    \label{fig:ViolinplotByMetricsCategory-FastJson}
\end{figure*}

\begin{figure*}
\includegraphics[scale=0.48]{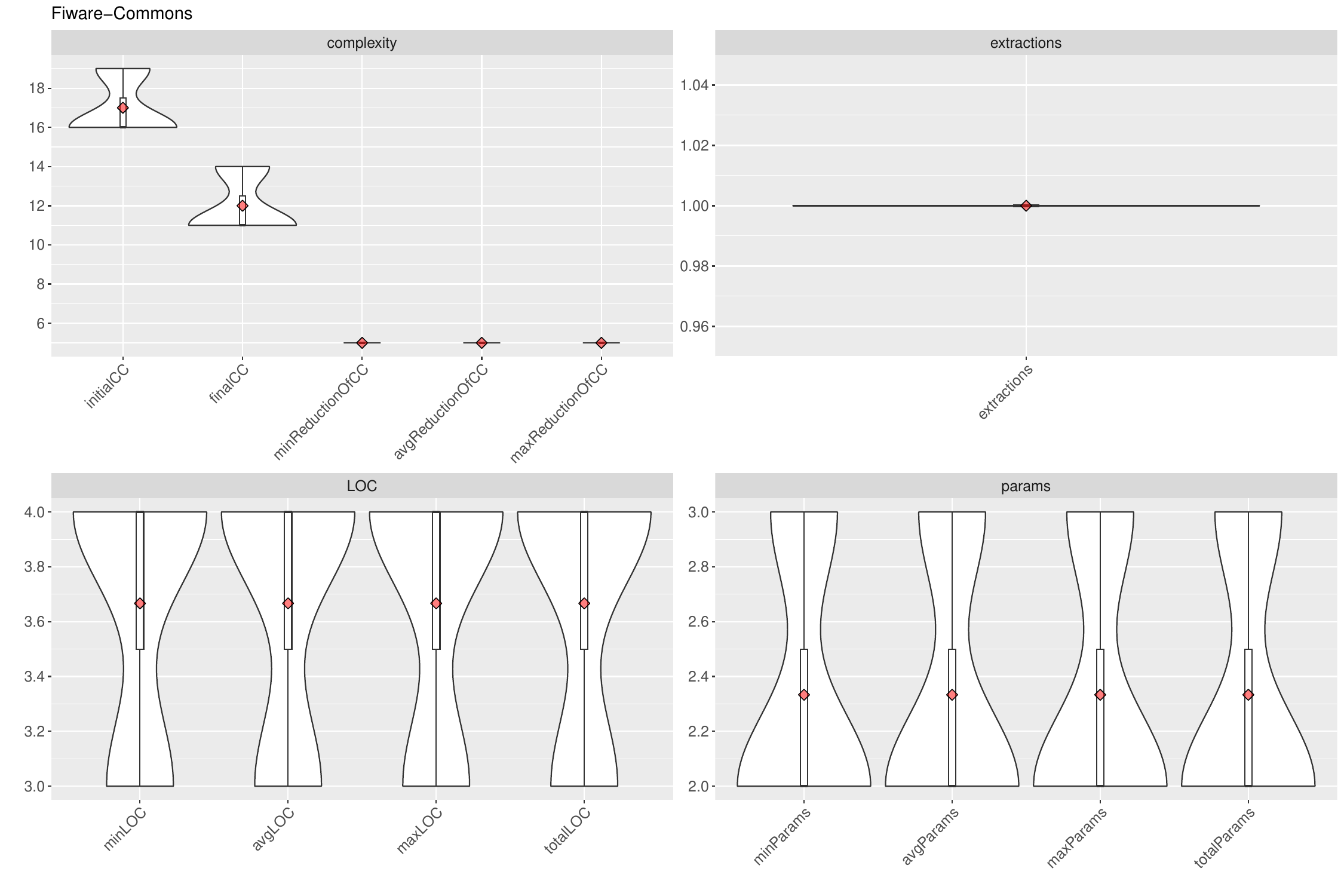}
    \caption{Summary statistics and the density of studied metrics for the 888 methods for which CPLEX found optimal solutions when using 300 seconds as stopping criterion though violin plots for the open-source project Fiware-Commons.}
    \label{fig:ViolinplotByMetricsCategory-Fiware-Commons}
\end{figure*}

\begin{figure*}
\includegraphics[scale=0.48]{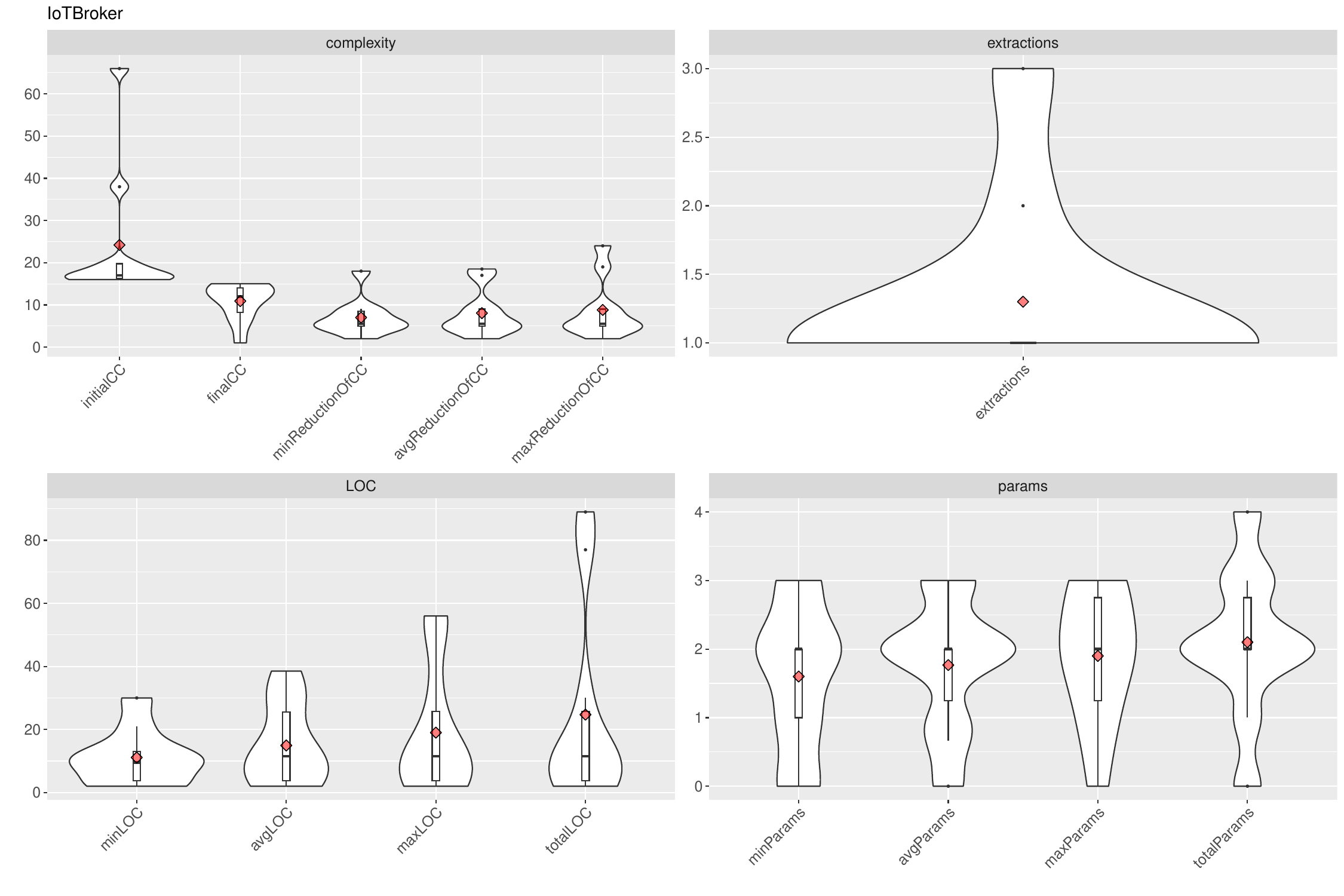}
    \caption{Summary statistics and the density of studied metrics for the 888 methods for which CPLEX found optimal solutions when using 300 seconds as stopping criterion though violin plots for the open-source project IoTBroker.}
    \label{fig:ViolinplotByMetricsCategory-IoTBroker}
\end{figure*}

\begin{figure*}
\includegraphics[scale=0.48]{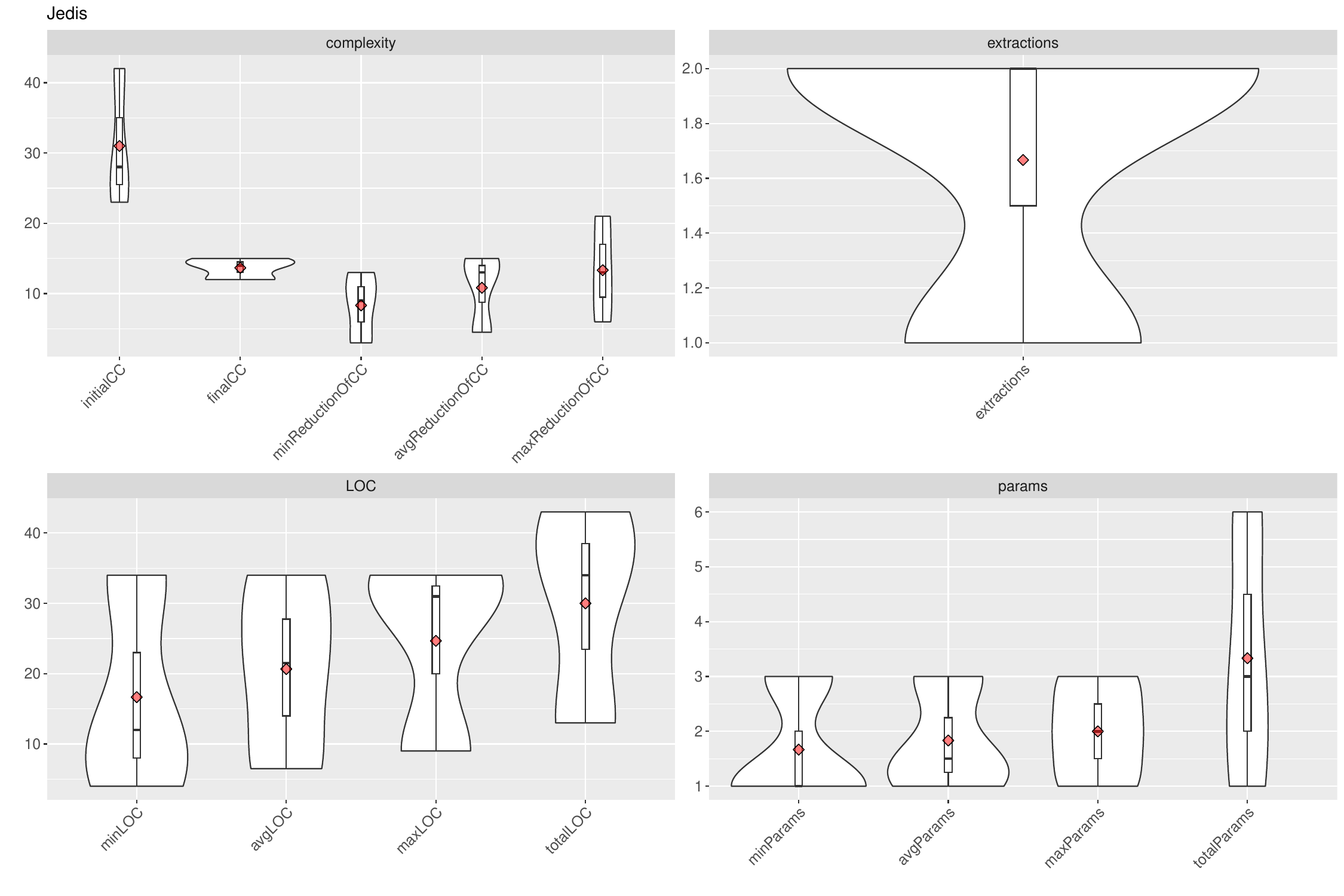}
    \caption{Summary statistics and the density of studied metrics for the 888 methods for which CPLEX found optimal solutions when using 300 seconds as stopping criterion though violin plots for the open-source project Jedis.}
    \label{fig:ViolinplotByMetricsCategory-Jedis}
\end{figure*}

\begin{figure*}
\includegraphics[scale=0.48]{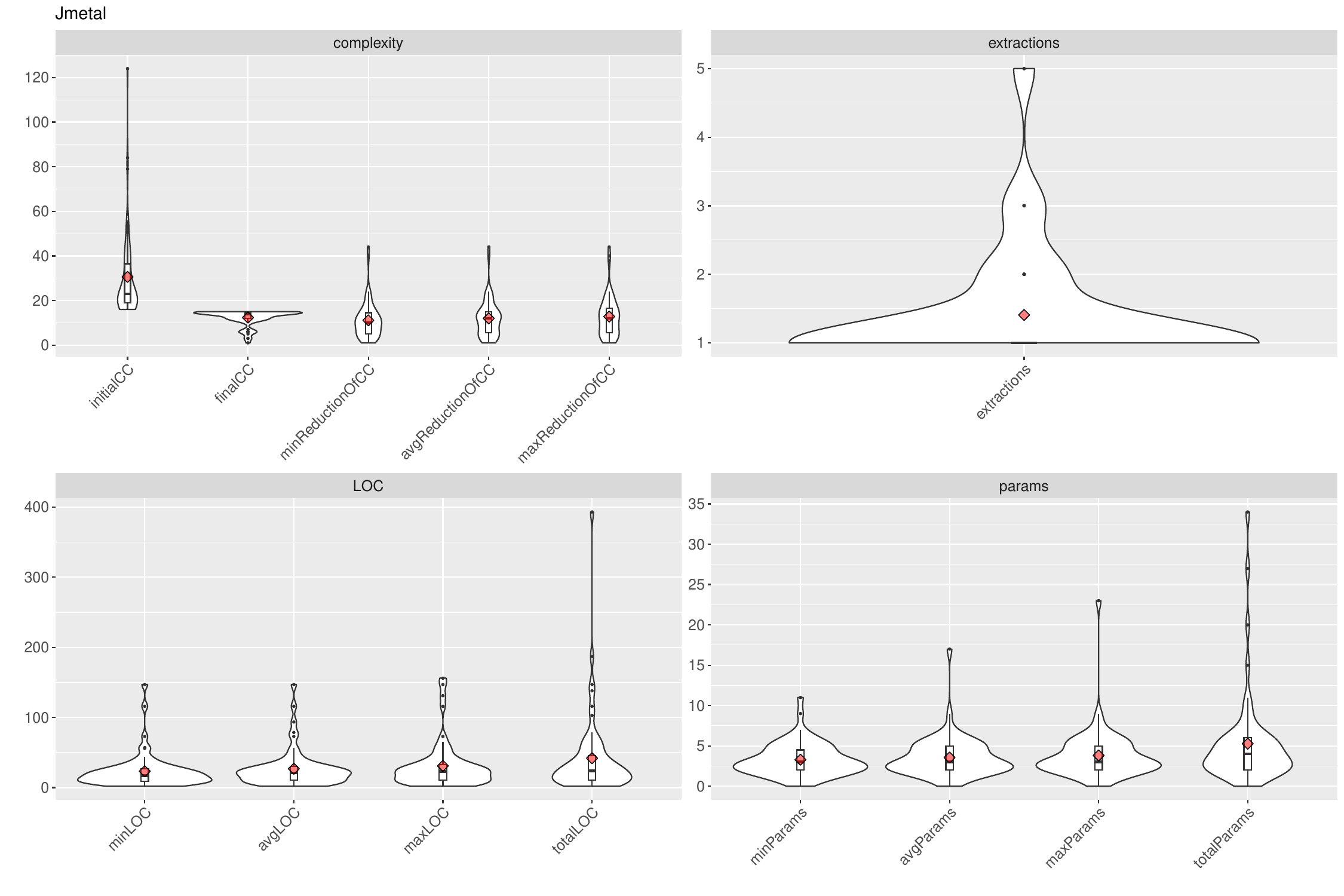}
    \caption{Summary statistics and the density of studied metrics for the 888 methods for which CPLEX found optimal solutions when using 300 seconds as stopping criterion though violin plots for the open-source project jMetal.}
    \label{fig:ViolinplotByMetricsCategory-Jmetal}
\end{figure*}

\begin{figure*}
 \includegraphics[scale=0.48]{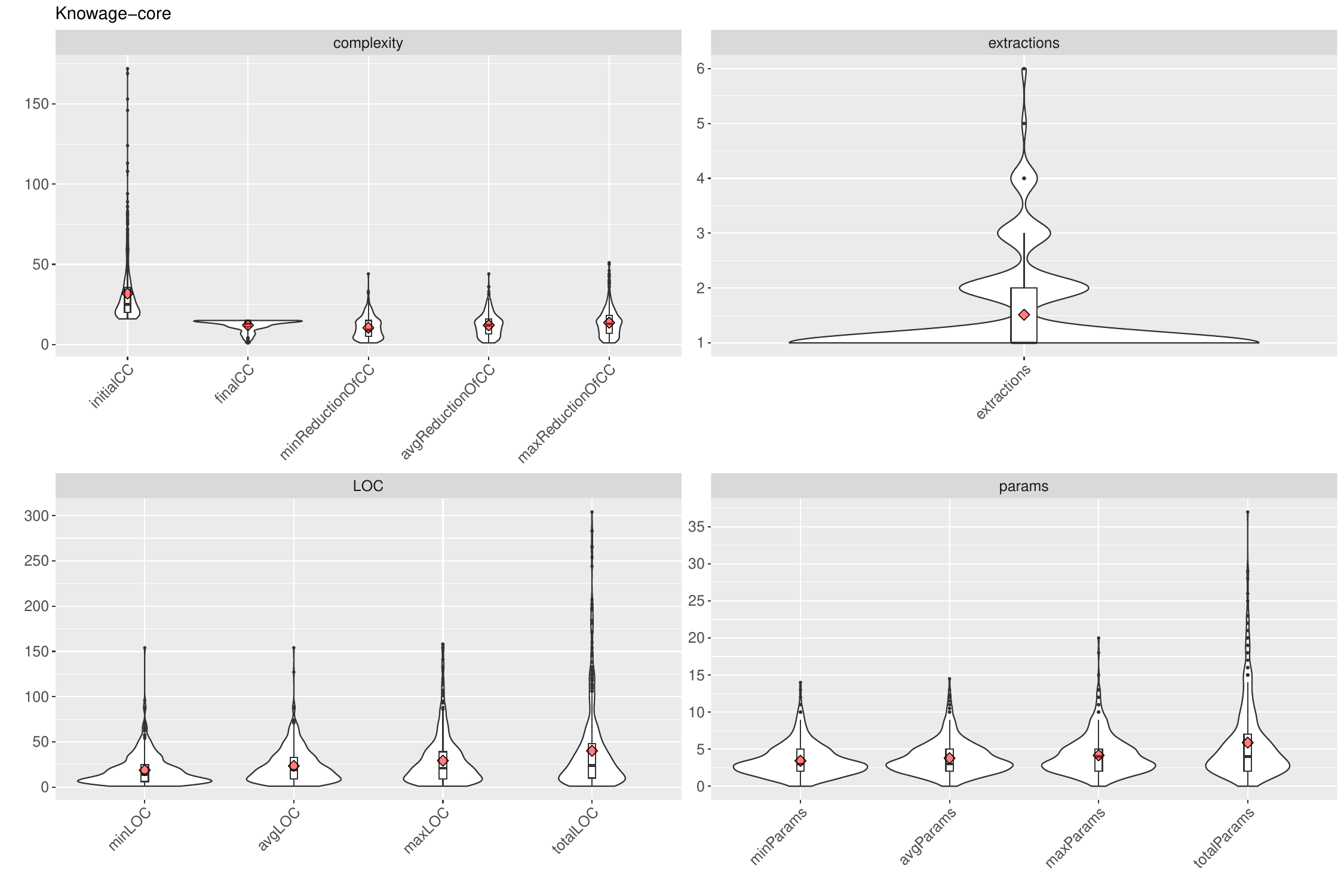}
    \caption{Summary statistics and the density of studied metrics for the 888 methods for which CPLEX found optimal solutions when using 300 seconds as stopping criterion though violin plots for the open-source project Knowage-core.}
    \label{fig:ViolinplotByMetricsCategory-Knowage-core}
\end{figure*}

\begin{figure*}
\includegraphics[scale=0.48]{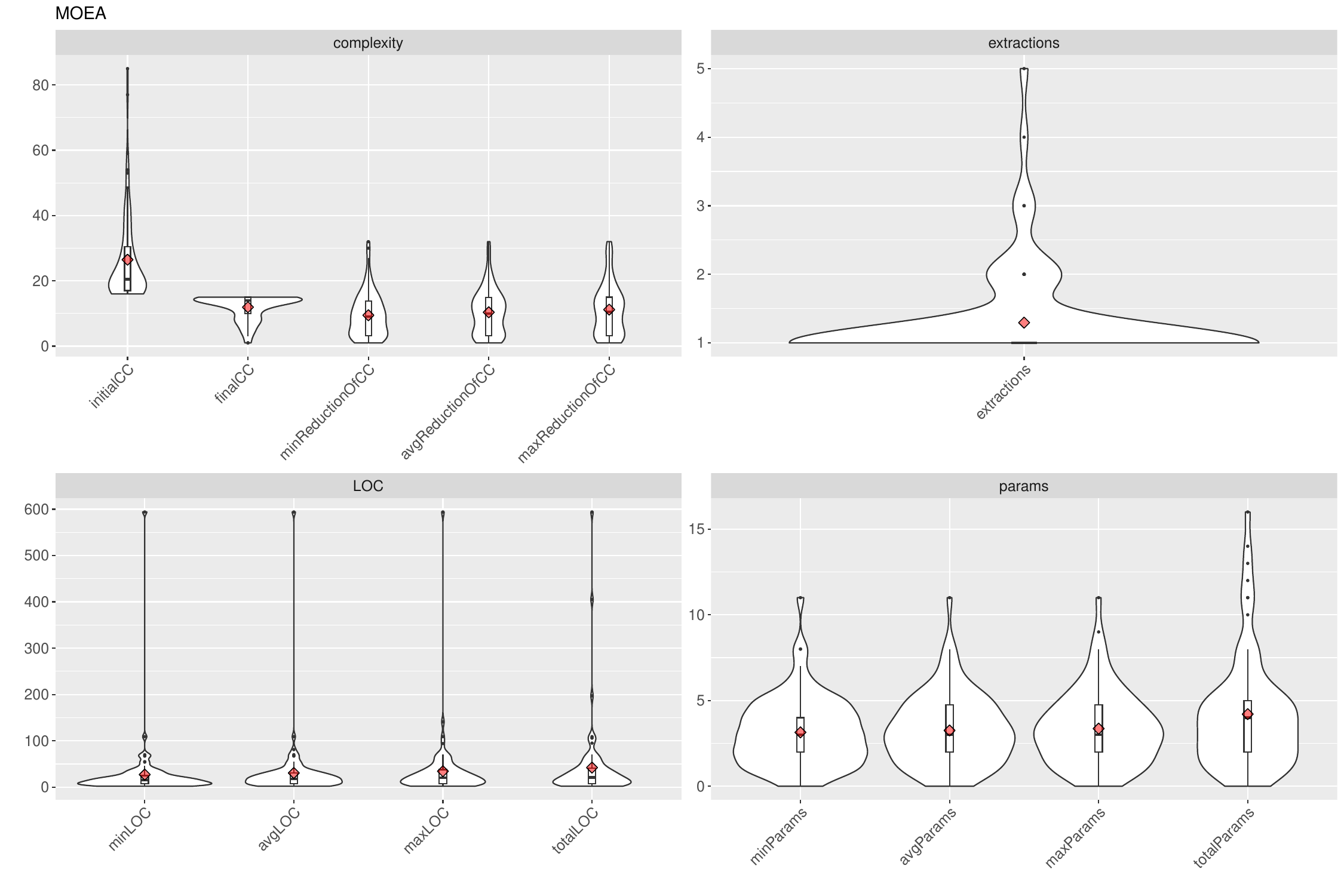}
    \caption{Summary statistics and the density of studied metrics for the 888 methods for which CPLEX found optimal solutions when using 300 seconds as stopping criterion though violin plots for the open-source project MOEA-framework.}
    \label{fig:ViolinplotByMetricsCategory-MOEA}
\end{figure*}

\begin{figure*}
\includegraphics[scale=0.48]{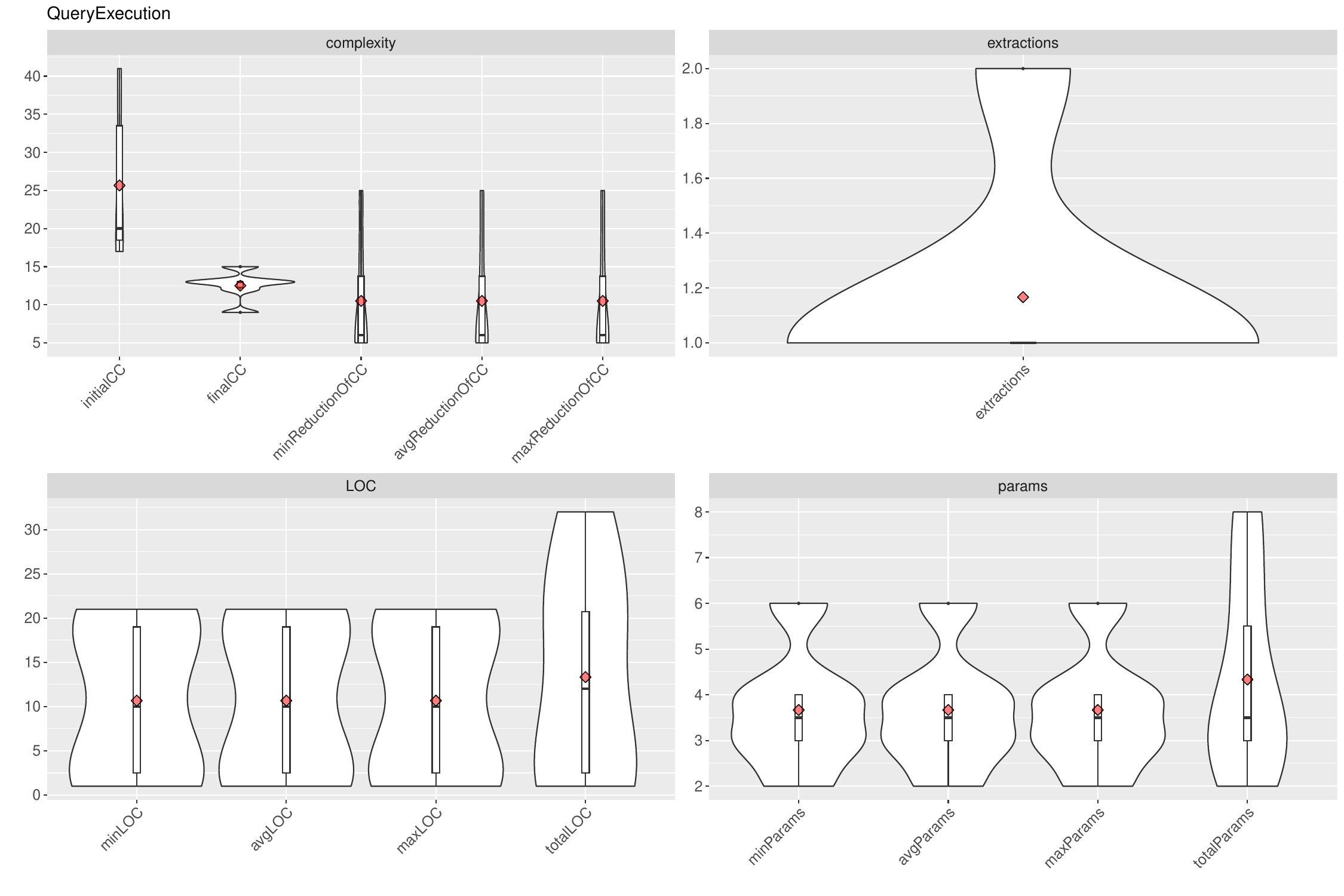}
    \caption{Summary statistics and the density of studied metrics for the 888 methods for which CPLEX found optimal solutions when using 300 seconds as stopping criterion though violin plots for the open-source project QueryExecution.}
    \label{fig:ViolinplotByMetricsCategory-QueryExecution}
\end{figure*}

\begin{figure*}
    \includegraphics[scale=0.45]{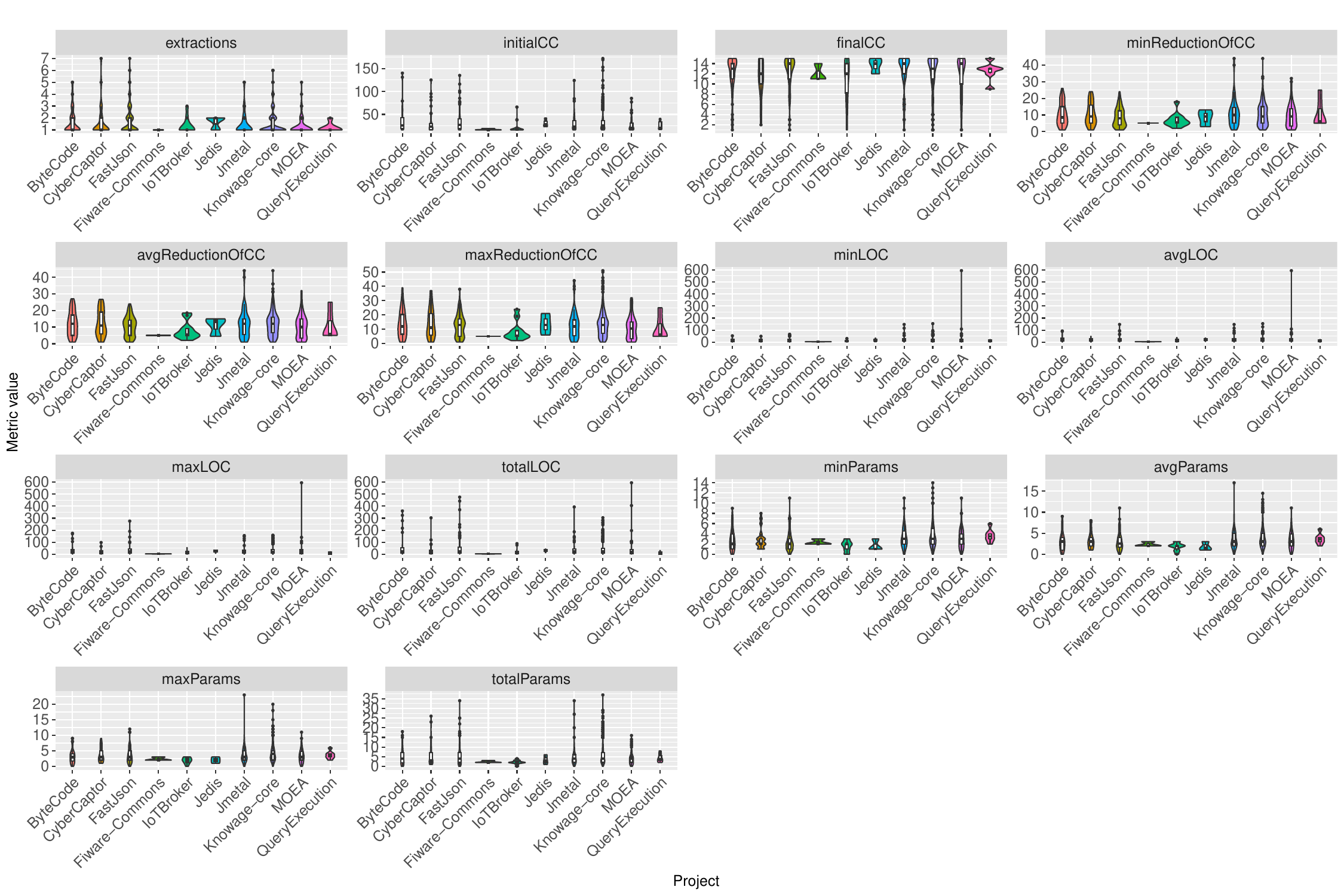}
    \caption{Summary statistics and the density by project and metric.}
    \label{fig:Violinplot}
\end{figure*}

\end{document}